\documentclass{aa}  
\usepackage{natbib}
\bibpunct{(}{)}{;}{a}{}{,}

\usepackage{graphicx}
\usepackage{txfonts}
\begin{document} 

\title{
Short-term variability and mass loss in Be stars
}
\subtitle
{
I. BRITE satellite photometry of $\eta$ and $\mu$ Centauri
\thanks
{
Based on data collected by the BRITE-Constellation satellite mission,
built, launched and operated thanks to support from the Austrian
Aeronautics and Space Agency and the University of Vienna, the
Canadian Space Agency (CSA), and the Foundation for Polish Science \&
Technology (FNiTP MNiSW) and National Science Centre (NCN).  Based 
in part also on observations collected at the European Organisation for 
Astronomical Research in the Southern Hemisphere under ESO programme 
093.D-0367(A).
}
}
\author{D.\,Baade\inst{1}
\and
Th.\,Rivinius\inst{2}
\and
A.\,Pigulski\inst{3}
\and 
A.C.\,Carciofi\inst{4}
\and
Ch.\,Martayan\inst{2}
\and 
A.F.J.\,Moffat\inst{5}
\and 
G.A.\,Wade\inst{6}
\and
W.W.\,Weiss\inst{7}
\and
J.\,Grunhut\inst{1}
\and
G.\,Handler\inst{8}
\and
R.\,Kuschnig\inst{9,}\inst{7}
\and
A.\,Mehner\inst{2}
\and
H.\,Pablo\inst{5}
\and
A.\,Popowicz\inst{10}
\and
S.\,Rucinski\inst{11}
\and
G.\,Whittaker\inst{11}
}

\institute{European Organisation for Astronomical Research in the 
Southern Hemisphere (ESO), Karl-Schwarzschild-Str.\,2, 
\newline
85748 Garching, Germany;   
\email{dbaade@eso.org}
\and
European Organisation for Astronomical Research in the 
Southern Hemisphere (ESO), Casilla 19001, Santiago 19, Chile 
\and
Astronomical Institute, Wroc{\l}aw University, Kopernika 11, 
51-622 Wroc{\l}aw, Poland
\and
Instituto de Astronomia, Geof{\' i}sica e Ci{\^ e}ncias Atmosf{\' e}ricas,
Universidade de S{\~ a}o Paulo, Rua do Mat{\~ a}o 1226, 
\newline
Cidade Universit{\' a}ria, 05508-900 S{\~ a}o Paulo, SP, Brazil
\and
D{\' e}partement de physique and Centre de Recherche en Astrophysique du 
Qu{\' e}bec (CRAQ), Universit{\' e} de Montr{\' e}al, C.P. 6128, 
Succ.\,Centre-Ville, Montr{\' e}al, Qu{\' e}bec, H3C 3J7, Canada 
\and
Department of Physics, Royal Military College of Canada, PO
Box 17000, Stn Forces, Kingston, Ontario K7K 7B4, Canada
\and
Institute of Astronomy, University of Vienna, Universit{\" a}tsring 1, 
1010 Vienna, Austria 
\and
Nicolaus Copernicus Astronomical Center, ul.\,Bartycka 18, 00-716 
Warsaw, Poland
\and
Department of Physics and Astronomy, University of British Columbia, 
Vancouver, BC V6T1Z1, Canada
\and
Institute of Automatic Control, Silesian University of Technology, 
Gliwice, Poland 
\and
Department of Astronomy \& Astrophysics, University of Toronto, 
50 St.\,George St, Toronto, Ontario, M5S 3H4, Canada 
}

\date{Received:  ; accepted:  }
 
\abstract
{ 
Empirical evidence for the involvement of nonradial pulsations (NRP's) in
the mass loss from Be stars ranges from (i) a singular case 
(\object{$\mu$ Cen}) of
repetitive mass ejections triggered by multi-mode beating to (ii) several
photometric reports about enormous numbers of pulsation modes popping 
up during outbursts and on to (iii) effective single-mode pulsators.
}
{
Develop a more detailed empirical description of the
star-to-disk mass transfer.  Check the hypothesis that spates of
transient nonradial pulsation modes accompany and even drive mass-loss
episodes. 
}
{ The BRITE Constellation of
  nanosatellites was used to obtain mmag photometry of the Be stars
  $\eta$ and \object{$\mu$ Cen}.  } 
{ In
  the low-inclination star \object{$\mu$ Cen}, light pollution by
  variable amounts of near-stellar matter prevented any new insights 
into the variability and other properties of the central star.  In the
  equator-on star \object{$\eta$ Cen}, BRITE photometry and {\sc Heros}
  echelle spectroscopy from the 1990s reveal an intricate clockwork of
  star-disk interactions.  The mass transfer is modulated with the
  frequency difference of two NRP modes and an amplitude three times
  as large as the amplitude sum of the two NRP modes.  This process
  feeds a high-amplitude circumstellar activity running with the
  incoherent and slightly lower so-called {\v S}tefl frequency.  The
  mass loss-modulation cycles are tightly coupled to variations in the
  value of the {\v S}tefl frequency and in its amplitude, albeit with
  strongly drifting phase differences.  } 
{ The observations are well
  described by the decomposition of the mass loss into a
  pulsation-related engine in the star and a viscosity-dominated
  engine in the circumstellar disk.  Arguments are developed that 
  large-scale gas-circulation flows occur at the interface.  The 
  propagation rates of these eddies manifest themselves as {\v
  S}tefl frequencies.  Bursts in power spectra during mass-loss events
  can be understood as the noise inherent to these gas flows. }


\titlerunning{BRITE satellite photometry of Be stars $\eta$ and \object{$\mu$ Cen}}
\authorrunning{D.\,Baade et al.}

   \maketitle
%

\section{Introduction} 

Be stars display one of the most ornate showcases of stellar physics
along the entire main sequence: extreme rotation, nonradial $g$- and
$p$-mode pulsation, outbursts, and high-speed winds.  And there are
the emission line-forming circumstellar disks that Be stars build as
the screens onto which they can project their activities. These
decretion disks are not just the antonym of accretion disks.  Instead
of powering high-energy jets to dispose of excess angular momentum,
they are struggling to reshuffle specific angular momentum such that
at least a fraction of the available matter can join the ranks of
Keplerian orbiters.  Once elevated, they slowly drift away, sometimes
in complex large-scale undulations.  Be stars avoid close companions
(unless they have swallowed them at an early moment) and do not
sustain large-scale magnetic fields.  The full Be-star saga is told by
\cite{2013A&ARv..21...69R} - as best as it can be done today.

Most notably, the physical process that enables Be stars to form
circumstellar Keplerian disks is not finally identified.  This
challenge decomposes into two parts.  The first one is to determine
the engine that accelerates stellar matter such that it either orbits
the star or drifts away from it.  A second engine is needed to
increase the specific angular momentum of this now circumstellar
matter so that it can reach Keplerian orbits with larger radii,
thereby forming a fully-developed, slowly expanding Keplerian disk.

Even at typically $\geq$70\% of the critical velocity \cite[Fig.\,9
in][]{2013A&ARv..21...69R}, rotation alone is not sufficient but most
probably a necessary part of it. Radiative winds are not a major
player because, in B-type main-sequence stars, they are very weak
\citep{1989MNRAS.241..721P, 2014A&A...564A..70K}. In Be stars, UV wind
lines are actually stronger than in B stars without circumstellar
disks \citep{1989ApJ...339..403G, 1989MNRAS.241..721P}.  But the
aspect dependence \citep{1987ApJ...320..376G} suggests that Be winds
are stronger because they are more easily launched in the zero-gravity
environment of the disk \citep[see also][]{2013A&ARv..21...69R}.

Since the first discoveries \citep{1982A&A...105...65B,
1982IAUS...98..181B} of nonradial pulsations (NRP's) in Be stars, it
has been hoped that they might provide the missing angular momentum
and energy to lift stellar matter into circumstellar orbits.  The
variability of the emission-line strength, which may drop to zero,
suggesting the complete dispersal of the disk
\citep[e.g.,][]{2010ApJ...709.1306W, 2012ApJ...744L..15C}, and
observations of discrete mass loss events in light curves
\citep{2009A&A...506...95H}, spectra \citep{1986ApJ...301L..61P,
1988A&A...198..211B}, and polarization \citep{1984ApJ...279..721H,
1984ApJ...287L..39G} have led to the notion that much, if not all, of
the star-to-disk mass-transfer process is episodic.

Already early on, indications were found that outbursts and changes in
the pulsation behavior of Be stars may be correlated
\citep{1982IAUS...98..181B, 1986PASP...98...35P}.  But in spite of
considerable and diverse efforts over 30+ years, only one case has
become known, in which NRP's are directly responsible for mass loss
episodes.  \cite{1998cvsw.conf..207R} found that during phases of
constructive superposition of the velocity field of the strongest mode
of \object{$\mu$ Cen} with that of the second-strongest or the
third-strongest mode, respectively, the H$\alpha$ line emission is
enhanced whereas the co-addition of the two weaker modes has no such
effect. Since the sum of the amplitudes $a_2$ and $a_3$ is smaller
than $a_1 + a_2$ and $a_1 + a_3$ while all three modes have the same
indices $\ell$ and $m$, the evidence seemed compelling that in
\object{$\mu$ Cen} the beating of NRP modes, with frequencies close to
2\,c/d, causes mass-loss events.

Once matter has managed to leave the star, the viscous decretion disk
(VDD) model \citep{1991MNRAS.250..432L} is generally acknowledged to
give the best current description of the build-up process of the disk
and its evolution \citep{2012ApJ...744L..15C,
2012ApJ...756..156H,2014ApJ...785...12H}.  It seems to be the solution
to the second part of the quest for the understanding of the Be
phenomenon, namely how Be stars develop Keplerian disks.

Other observational studies have reported somewhat different relations
but have drawn essentially the same conclusion, namely that (some) Be stars
owe their disks to (the combination of rapid rotation and) nonradial
pulsation.  These examples will be discussed below after some
additional empirical arguments have been developed so that they enable
a different interpretation, inspired by the variability of $\eta$ and
\object{$\mu$ Cen} and first anticipated by
\cite{2013ASSP...31..253R}. 

The complexity of the matter is highlighted by the fully negative
results to date of searches for \object{$\mu$ Cen} analogs.  Moreover,
28 $\omega$ CMa is perhaps the spectroscopically best-established
single-mode pulsator among Be stars \citep{2003A&A...411..167S}.  And
yet it shows highly variable H$\alpha$ emission-line strength and,
therefore, mass-loss rate.  However, at 7-9 years
\citep{2003A&A...402..253S, 2012ApJ...744L..15C}, the repetition time
of its outbursts is much more than an order of magnitude longer than
in \object{$\mu$ Cen}.

On the other hand, \object{$\mu$ Cen} is one of the closest Be stars,
and it would be odd if it were a singular case.  Perhaps, a rich
source of \object{$\mu$ Cen} analogues is the MACHO database
\citep{2003ASPC..292...97K}.  Many of the false-alarm lensing events
therein were due to outbursts of Be stars that satisfied the lensing
selection criteria: roughly color-neutral brightening with an
approximately symmetric light curve.  In some cases, the outbursts
repeated semi-regularly, which might be the signature of multi-mode
beating.  The OGLE database could harbour a similar bonanza
\citep{2002A&A...393..887M}, and the same may hold for the ASAS
database \citep{2005AcA....55..275P} as well.

In short, Be stars require one engine to expell matter and a second
one to arrange the ejecta in a slowly outflowing Keplerian disk.
Detailed comparisons of model calculations and observations have
established the VDD model as the basis from which to further explore
the disk properties.  The broad acceptance of the NRP hypothesis for
the inner engine rests on much circumstantial evidence but also on the
lack of other ideas.  The interface between these two engines is largely 
unexplained territory.  

One of the premises, on which the BRITE project is built, is that for
bright stars extensive series of high-quality spectra can be obtained
or are already available.  The present study will play this card for
$\eta$ and \object{$\mu$ Cen}, which have been intensively monitored
with the {\sc Heros} spectrograph \citep{1997A&A...320..273K,
1997A&A...328..219S} in the 1990s.  It is the first paper in a series
of studies revisiting the variability of Be stars with space
photometers.  The next one (Rivinius et al., in preparation; hereafter
"Paper II") re-analyses {\it Kepler} observations.  The results complement,
and partly extend, those developed in this work.

Before presenting the BRITE observations (Sect.\,3), the method used
for their time-series analysis (Sect.\,4), the results (Sect.\,5),
their analysis (Sect.\,6) and discussion (Sect.\,7), and finally the
conclusions (Sect.\,8), it is useful to describe in the next section a
process that has not so far required and, therefore found, much
attention.  But for the understanding of the observations presented in
this paper it is of central importance.

\begin{table}
\caption{Characteristic frequencies (in c/d) of Be stars known to exhibit 
{\v S}tefl frequencies (N/A = not available; errors of stellar rotation 
and maximal Kepler frequencies may be of order 10\%, errors of {\v S}tefl and 
nearest NRP frequencies are below 0.01\,c/d.).  References and comments 
are provided in Sect.\,\ref{relrot}.}
\label{freqtab}      
\centering                        
\begin{tabular}{c c c c c c}        
\hline\hline                
Star            &   \multicolumn{4}{c}{Frequency type}           & NRP         \\  
                & Stellar   & Maxim.   &  {\v S}tefl & Nearest  & mode(s)      \\ 
                & rotat.  & Kepler    &             & NRP      &   \\
\hline                       
\object{$\mu$ Cen}       &   2.1     &   2.7     &   1.6       &  1.94    &      mult.      \\
\object{$\eta$ Cen}      &   1.7     &   2.2     &   1.56      &  1.73    &      mult.      \\
\object{\object{28 $\omega$ CMa}} &   0.9     &   1.1     &   0.67      &  0.73    &      single      \\
\object{\object{$\kappa$ CMa}}    &   N/A     &   N/A     &   1.62      &  1.83    &    N/A      \\
\object{Achernar}        &   0.64    &  0.77     &   0.73      &  0.78    &      single      \\
\hline                                   
\end{tabular}
\end{table}

\section{{\v S}tefl frequencies}
\subsection{Spectroscopic signatures}

Nearly 20 years ago, \cite{1998ASPC..135..348S} discovered that,
during outbursts (periods of temporarily enhanced H$\alpha$ line
emission), the Be stars \object{28 $\omega$ CMa} \citep[see also][ B2
IV-Ve]{2000ASPC..214..240S} and \object{$\mu$ Cen} (B2 Ve) develop
variability in the profiles of spectral lines normally formed above
the photosphere (e.g., Mg\,II 448.1, Si\,II 634.7, etc.\ in
absorption, the violet-to-red emission-peak ratio, V/R, of Fe\,II
531.6, and the mode of Balmer emission-line profiles).  In agreement
with this, the variability was confined to projected velocities above
the equatorial level.  (This agreement does not necessarily imply
confirmation because the structures involved may have their own
associated velocity field [as in pulsations].)  The super-photospheric
location was further observationally supported by
\cite{1998A&A...333..125R}, who found that the violet-to-red
emission-peak ratio, $V/R$, of double-peaked emission lines in
\object{$\mu$ Cen} is part of this variability also.

In both stars, these temporary so-called {\v S}tefl frequencies are
about 10-20\% lower than that of a nearby strong stellar
frequency, which the underlying line-profile variability unambiguously
identified as due to nonradial pulsation.

\cite{1998ASPC..135..348S} suspected \object{$\eta$ Cen} to
be a third case.  This was confirmed by \cite{2003A&A...411..229R} as
described in more detail in Sect.\,\ref{etaOLDspec}. An interesting
difference with respect to the other two stars is that in
\object{$\eta$ Cen} the {\v S}tefl frequency seems to be permanent and
of much larger amplitude than the stellar pulsations.
\cite{2003A&A...411..229R} also added \object{$\kappa$ CMa} as a
fourth Be star with transient frequencies.  Using several lines they
not only illustrated (their Fig.\,15) in more detail that the velocity
range of the features substantially exceeds $\pm v \sin i$ but
also that quasi-periodic line profile-crossing features are associated
with this variability.

The two discovery publications received hardly any citations, and even
the authors themselves only proposed \citep{1998A&A...333..125R} the
very qualitative idea of an ejected cloud with an orbit that has not
yet been circularized before merging with the disk.  It is not clear
how this notion can account for the permanent presence of a {\v S}tefl
frequency in \object{$\eta$ Cen} (see also the next subsection for the
somewhat similar case of \object{Achernar}).  Therefore, {\v S}tefl
frequencies mark one of the largest mostly comprehension-free domains
of observational knowledge of Be stars.  They appear to arise from the
interface between star and disk, the engine generating them is
probably fed by the mass-loss process, and the latter may involve 
stellar NRP modes of slightly higher frequency.  Compared to stellar
oscillations, {\v S}tefl frequencies may carry relatively little
global information about the central star.

\subsection{Photometric signatures}

Before BRITE, long series of high-quality photometric observations
were not available for any of the stars with spectroscopic {\v S}tefl
frequencies.  Without spectroscopic diagnostic support, these
frequencies can be a bit treacherous because they may be mistaken for
stellar pulsations.  However, this extra difficulty makes the
identification of {\v S}tefl frequencies in photometric data by no
means impossible.

The best test case at hand are the observations of
\object{$\alpha$ Eri} (Achernar; B4) with the {\it Solar Mass
Ejection Imager (SMEI)}.  \cite{2011MNRAS.411..162G} have published a
very elaborate analysis, which concludes that there are only two
significant frequencies.  F1 = 0.775 c/d is fairly constant over
nearly 5 years while F2 = 0.725 c/d exhibits significant shifts in
frequency and phase, which moreover are associated with apparent
overall brightness variations. Both variabilities have time-dependent
amplitudes.  With a factor of 8, the amplitude variation of F2 is the
strongest, which even dropped below detectability.

The nature of the {\it SMEI} observations does not readily permit
detection of long-term brightness variations with confidence.
Moreover, \object{$\alpha$ Eri} is observed at a large inclination
angle but not equator-on.  Model calulations by
\cite{2012ApJ...756..156H} (their Fig.\ 13) suggest that in such a
case the response in optical light to the ejection of matter is rather
minimal.  Therefore, the annual {\it SMEI} light curves are not really
sufficient to infer the occurrence of an outburst at the onset in 2004
of the various anomalies described above.  But
\cite{2013A&A...559L...4R} report a strengthening of the H$\alpha$
equivalent width at the end of 2004, which is the signature of an
increased amount of matter close to the central star.

Figure 2 of \cite{2011MNRAS.411..162G} is of special interest:
Although it covers 30 d, it exhibits no indication of the nominal beat
period of about 20 d for F1 and F2.  This implies quite immediately
that one or both of the two variations are not phase coherent.

Since the incoherent frequency F2 is smaller than the coherent F1, the
analogy to the spectroscopic examples of the previous subsection
suggests that F1 is a stellar pulsation while F2 would be a
circumstellar {\v S}tefl frequency.

\subsection{Relation to rotation and revolution} 
\label{relrot}

Two frequencies warrant comparison with the {\v S}tefl frequencies:
that of the stellar rotation and that corresponding to the innermost
Keplerian orbit possible within a disk.

The {\v S}tefl frequency of \object{$\mu$ Cen} as reported in the
discovery paper \citep{1998ASPC..135..348S} is 1.59\,c/d;
\cite{2003A&A...411..229R} later found it closer to 1.61\,c/d.  The
nearest pulsation frequency is 1.94\,c/d \citep{1998A&A...336..177R}.
The work of \cite{2001A&A...369.1058R} leads to a stellar rotational
frequency, $f_{\rm rot}$, of 2.1\,c/d and a maximal Keplerian
frequency, $f_{\rm Kepler}$, of 2.7\,c/d.

For \object{$\eta$ Cen}, \cite{2006A&A...459..137R} list a projected
rotational velocity of 350\,km\,s$^{-1}$ and a fractional critical rotation of
0.79.  Because the star is a shell star, sin\,$i$ can be assumed to be
$\geq$0.95. $\eta$ and \object{$\mu$ Cen} have similar spectral types
and colors \citep{2002yCat.2237....0D} and the difference in apparent
brightness \citep[B magnitudes: 2.1 and 3.3; ][]{2002yCat.2237....0D}
is consistent with the difference in parallax \citep[10.67 and
6.45\,mas; ][]{1997A&A...323L..49P}.  Therefore, the radius,
4.2\,$R_{\odot}$, determined by \cite{2001A&A...369.1058R} for $\mu$
Cen is also adopted for \object{$\eta$ Cen}. If also the masses are
equal, $f_{\rm rot}$ = 1.7\,c/d and $f_{\rm Kepler}$ = 2.2\,c/d.  The
{\v S}tefl frequency is 1.56 c/d \citep{2003A&A...411..229R}, and the
same authors found the nearest NRP frequency of \object{$\eta$ Cen} at
1.73\,c/d. 

The {\v S}tefl frequency of \object{28 $\omega$ CMa} is 0.67\,c/d
\citep{1998ASPC..135..348S}. \cite{1982A&A...105...65B} determined the
nearest neighbouring stellar frequency at 0.73\,c/d.  The work of
\cite{2003A&A...411..181M} led to $f_{\rm rot}$ = 0.9\,c/d and
$f_{\rm Kepler}$ = 1.1\,c/d.

\cite{2003A&A...411..229R} noted that \object{$\kappa$ CMa} is one of
the few objects in their sample of 27 Be stars not showing line
profile-variability with the characteristics of quadrupole modes.
They determined a period of the continuous variability of 1.825\,c/d
while the transient frequency amounted to 1.621\,c/d.

For \object{Achernar}, \cite{2014A&A...569A..10D} published $f_{\rm
rot}$ = 0.644\,c/d.  From this and the other parameters provided in
that study, the value of $f_{\rm Kepler}$ was determined as
0.769\,c/d.  The photometric, and probable {\v S}tefl, frequency F2
reported by \cite{2011MNRAS.411..162G} is 0.725 c/d.  The stellar F1
of \cite{2011MNRAS.411..162G} occurs at 0.775\,c/d, and it seems
remarkable that for all practical purposes it is identical to the
Kepler frequency.

For easier comparison, all frequencies are listed in
Table\,\ref{freqtab}.

\begin{table*}
\caption{Overview of BRITE observations}
\label{overviewtab}
\centering
\begin{tabular}{l c c c r r c}
\hline\hline
Satellite name  &  Passband  &  Orbital period  &  Contiguous time  &  JD start  &  JD end &  Range in CCDT \\
                &            &  [min]           &  [min]            &  \multicolumn{2}{c}{-2,450,000} &  [\degr C]  \\
\hline
BRITE-Austria   &  blue      &  100.4           &  6--16            &   6756.287  &  6887.910 &  17--38  \\
BRITE-Lem       &  blue      &  99.6            &  10--12           &   6820.808  &  6857.456 &  12--17  \\
UniBRITE        &  red       &  100.4           &  9--14            &   6742.186  &  6887.471 &  16--40  \\
BRITE-Toronto   &  red       &  98.2            &  17               &   6835.817  &  6841.783 &  13--16  \\
\hline
\end{tabular}
\tablefoot{'Contiguous time' denotes the typical time interval per orbit, 
during which exposures were made.}
\end{table*}

\section{Observations}

\subsection{BRITE Constellation}

The new observations presented and discussed in this paper are
photometric monitoring data obtained with the BRITE
Constellation.  It consists of a cluster of five nearly identical
nanosatellites and is described in detail by
\cite{2014PASP..126..573W} and Pablo et al.\,(in preparation).  The
satellites have an aperture of 3 cm and no moving parts.  Fixed
filters make three of them red-sensitive and the other two
blue-sensitive.  The roughly rectangular transmission curves define
wavelength passbands of 390-460\,nm and 550-700\,nm, respectively.
The field of view is 20\,x\,24\,deg$^{2}$.  In order to achieve
sub-mmag sensitivity, the CCD detectors are not in the focal plane,
which avoids saturation and reduces the impact of detector blemishes.
Data for up to $\sim$20-30 stars are simultaneously extracted and
downlinked to the ground stations.  The orbital periods are close to
100 minutes, enabling continuous observations for about 5-20 minutes. 
In these intervals, one-second exposures were made every 15-25 seconds.  

The observations used for this study were acquired in 2014 April-July
with satellites {\it BRITE-Austria} and {\it BRITE-Lem}
(blue-sensitive) and {\it Uni-BRITE} and {\it BRITE-Toronto}
(red-sensitive).  {\it BRITE-Austria} \& {\it Uni-BRITE} as well as
{\it BRITE-Lem} \& {\it BRITE-Toronto} were simultaneously pointed at
the BRITE Centaurus 2014 field, which contained both $\eta$ and
\object{$\mu$ Cen}.

\subsection{Raw database}

BRITE investigators are provided with ASCII tables containing the
following information (see also Pigulski et al.\ [submitted to A\&A] 
and Pablo et al.\ [in preparation]):
\begin{list}{$\bullet$}{\partopsep=0mm\topsep=0mm\itemsep=0mm}
\item
Flux: in instrumental units and for 1-s integrations (BRITE does not observe
photometric standards)
\item
Julian Date
\item
HJD: Heliocentric Julian Date
\item
XCEN:  X coordinate (in CCD pixels) of the stellar image
\item
YCEN:  Y coordinate (in CCD pixels) of the stellar image
\item
CCDT:  CCD temperature (in \degr C)
\end{list}

\noindent
The methods used to build these data packages are elaborated by 
Pablo et al.\,(in prep.) and Popowicz et al.\,(in prep.).  
Key challenges include:
\begin{list}{$\bullet$}{\partopsep=0mm\topsep=0mm\itemsep=0mm}
\item
The very complex optical point spread function (PSF) varies strongly
across the field.  Simple PSF fitting is not possible, and aperture
photometry is used.
\item
The pointing stability of the satellites is not perfect, and
occasionally stars drift out of the aperture, over which the signal is
extracted.
\item
The CCDs are not radiation hard and are deteriorating in quality and 
the number of bad pixels is increasing with time.
\item
The CCDs are not actively cooled.  Their sensitivity is temperature
dependent.
\end{list}

\noindent
Table \ref{overviewtab} provides a summarizing overview of
the contents of the ASCII tables with the raw data available for this
investigation of $\eta$ and \object{$\mu$ Cen}. 
 
Prior to any time-series analysis, some post-processing of the raw
data is needed to address the issues mentioned.

\subsection{Data processing}
 
This data conditioning consisted of the following steps applied
separately to each data string (one for each of the four BRITE
satellites used):

\begin{enumerate} 
\item The variability of the flux was
determined for each orbit.  Orbits with >2.5-$\sigma$ deviation from
the mean orbital variability were fully discarded.  
\item Within each
remaining orbit, individual 1-second data points differing by more
than 2.5 $\sigma$ from the orbital mean were deleted.  
\item In each
orbit, the remaining data were averaged to 1 or 2 bins of equal width
in time so that, in the case of such splitting, the number of averaged
1-second datapoints was at least 45.  
\item The so binned data were
plotted versus CCDT, XCEN, YCEN, and time and simultaneously
displayed.  In an interactive iterative procedure, the respective
strongest trend was fitted with a first- (mostly) or second- (rarely)
order polynomial, which was subtracted.  The plots were updated and,
if necessary, any remaining trend was corrected for.  Typically, one 
or two such regression analyses were performed and applied. 
\end{enumerate}
A consequence of these corrections is that the mean magnitudes are 
about zero.  

The data strings were combined to single datasets,
separately for each passband and star.  In some cases, a small
constant offset was applied to individual data strings to further
reduce any large-scale structure in the light curves.

Note that this conditioning of the raw data for the subsequent
time-series analysis is biased to enabling the detection of periods
significantly shorter than the length in time of the data strings.  It
does not introduce spurious variability of this kind but may distort
the light curve on longer timescales.

The final datasets prepared for time-series analysis are
characterized in Tables \ref{etadata} and \ref{mudata}.

Pigulski et al. (submitted to A\&A) provide a very useful and much
more comprehensive (but not fundamentally different) description of
how BRITE data can be post-processed.

\begin{table}
\caption{BRITE database for \object{$\eta$ Cen} (B2 Vnpe; V $\sim$2.3 mag; HR 5440; HD 127973)}
\label{etadata}      
\centering                        
\begin{tabular}{l r r r}        
\hline\hline              
Satellite      & No.\,of 1-s  &  No.\,of          &  No.\,of datapoints  \\   
               & exposures    &  orbits used      &  for TSA \\ 
\hline                                    
BRITE-Austria  &  39,801      &  1016             &  1095                \\
BRITE-Lem      &   3,917      &  129              &  117                 \\
UniBRITE       &  64,807      &  1213             &  1918                \\
BRITE-Toronto  &   4,935      &  54               &  95                  \\
\hline                                   
\end{tabular}
\end{table}

\begin{table}
\caption{BRITE database for \object{$\mu$ Cen} (B2 Ve; V $\sim$3.4 mag; HR 5193; HD 120324)}            
\label{mudata}      
\centering                        
\begin{tabular}{l r r r}        
\hline\hline                
Satellite      & No.\,of 1-s  &  No.\,of          &  No.\,of datapoints  \\   
               & exposures    &  orbits used      &  for TSA \\ 
\hline                       
BRITE-Austria  &  39,857      &  1017             &  1149                \\
BRITE-Lem      &   3,960      &   124             &   125                \\
UniBRITE       &  65,944      &  1165             &  1846                \\
BRITE-Toronto  &   4,953      &    53             &   102                \\
\hline                                   
\end{tabular}
\end{table}

\section{Time-series analysis}

\subsection{Method}
\label{method}

\begin{figure}
\includegraphics[width=6.2cm,angle=-90]{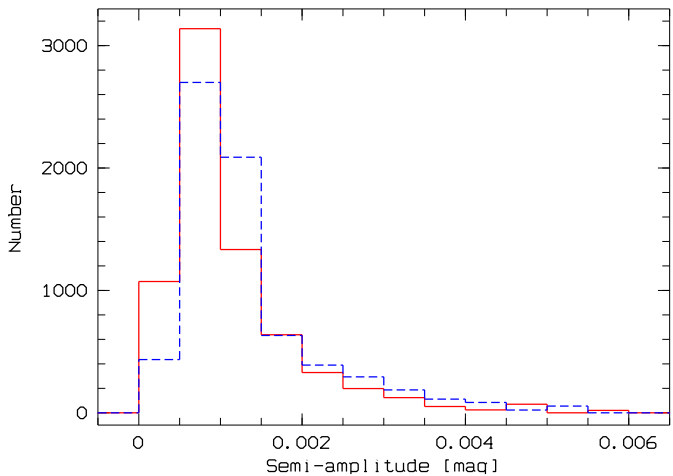}
\caption{
Histogram of the amplitudes of sine curves fitted to the blue 
(dotted line) and red (solid line) light curve of \object{$\eta$ Cen} 
(pre-whitened for the circumstellar 1.556\,c/d variation).  
The frequency range is 1\,c/d to 8\,c/d with a step of 0.001\,c/d.
}
\label{histogram}
\end{figure}

\begin{figure} 
\includegraphics[width=6.2cm,angle=-90]{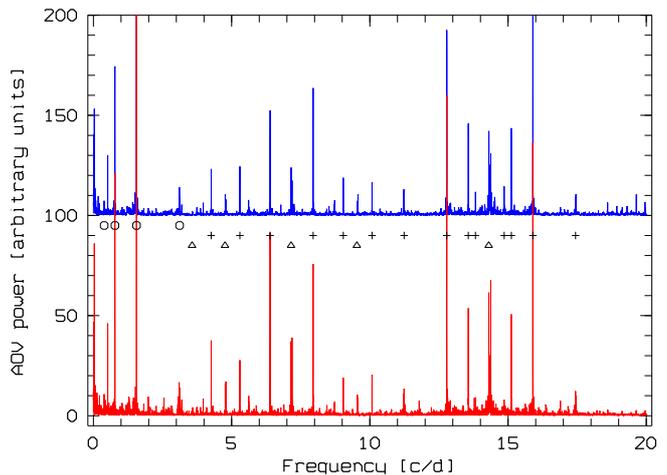} 
\caption
{ 
AOV spectra of \object{$\eta$ Cen}; that for the blue data is vertically
offset.  Because the orbital frequencies of the two satellites are
very similar, the AOV spectra look similar as well, and the two
spectra also illustrate the window function.  The orbital frequency,
$f_o$, at $\sim$14.4\,c/d and its subharmonics are marked by
triangles, the 1.5562\,c/d {\v S}tefl frequency and its harmonics and
sub-harmonics by circles, and features due to combinations of the two
by crosses. For stellar pulsations, $f_i$, which typically have
amplitudes more than an order of magnitude smaller than the {\v S}tefl
frequency, $f_o-f_i$ are the strongest aliases.  For $f_i<f_o/2$
($\sim$7.2\,c/d), genuine frequencies and aliases with $f_o$ do not
overlap.  Since the Nyquist frequency is about $f_o$/2, the range
below it, $f_i < f_o$/2, is not seriously contaminated by orbital
aliases.  
} 
\label{etaAOV20} 
\end{figure}

Time-series analyses (TSA) in the range 0-20\,c/d were carried out
with ESO-MIDAS \citep{2003ASSL..285...89B} context TSA, using the
Analysis of Variance \citep[AOV; ][]{1989MNRAS.241..153S} method.  For
comparison, conventional power spectra were also calculated.  There
were no major differences between the results.

Some periodic large-amplitude variations as described below for the
two stars had to be removed to enable searches for weaker variations.
This pre-whitening was performed by folding the data with the period
in question, binning them to 0.02 in phase, and subtracting from each
data point the average value of its home bin.

To get a first overview of the properties of the data as described
above, sine curves were fitted at every frequency between 1 and 8\,c/d
with a step of 0.001\,c/d. From the histogram of the fitted
semi-amplitudes (cf.\ Fig.\,\ref{histogram}), an initial quantitative
indicator of the significance threshold for periodic variations can be
deduced.  If it is required that for an AOV feature to be considered
significant its strength exceed that at the peak of the histogram plus
three times the $\sigma$ of the distribution, thresholds for the blue
and red data of \object{$\eta$ Cen} are 4.0\,mmag and 3.6\,mmag,
respectively, in the range 1-8\,c/d.  The corresponding values for
\object{$\mu$ Cen} are 3.3\,mmag and 6.4\,mmag.  Analysis of constant,
or nearly constant, Be stars shows (Baade et al., in prep.) that such
estimates of the performance of BRITE-Constellation are
over-conservative.  But in $\mu$ and $\eta$ Cen there is also stellar
noise that increases the detection thresholds.

For lower-amplitude variations, the often-adopted method of recursive
pre-whitening was not applied.  This is afforded by the fact that the
strongest aliases, $f_o - f_i$, of intrinsic frequencies $f_i$ arise
from the orbital frequency $f_o \approx 14.4$\,c/d, which is
analogous to 1\,c/d in ground-based data.  In BRITE data, 1-c/d
aliases are mostly weak but not completely absent due to the daily
variations of the solar illumination.  In Fig.\,\ref{etaAOV20} they
are not visible.  For frequencies $f_i < f_{o}/2$, aliases of $f_o$
cannot be confused with the intrinsic frequencies $f_i$ (this relation
is illustrated in Fig.\,\ref{etaAOV20}). It is, then, convenient that
the effective Nyquist frequency is $f_{o}$/2 (there is mostly one data
point per orbit) and the bulk of pulsation frequencies of Be stars
occur in the range 0.5-10\,c/d.  

Every AOV feature that by its strength relative to its surroundings
seemed to be possibly significant was checked by fitting a sine curve
with the respective frequency.  The analysis of the results suggested
as an empirical (necessary but not sufficient) criterion that at least
two of $f_i$/2, $f_o - f_i$, and $f_o + f_i$ should be clearly present
in the AOV spectrum for $f_i$ to be considered significant.  This
criterion provides a stronger filter than analysis of single AOV
features only.  In a first step, searches for frequencies were also
carried out in the product of the blue and red AOV spectra, which has
a higher contrast. This method is very effective but has the drawback
of weakening the signature of low-amplitude variations and those with
strong colour dependencies.

The final results stated in the text and in Table \ref{etafreq} were
obtained by fitting sine curves, starting with the frequencies found
by AOV. 

\begin{figure}
\includegraphics[width=6.2cm,angle=-90]{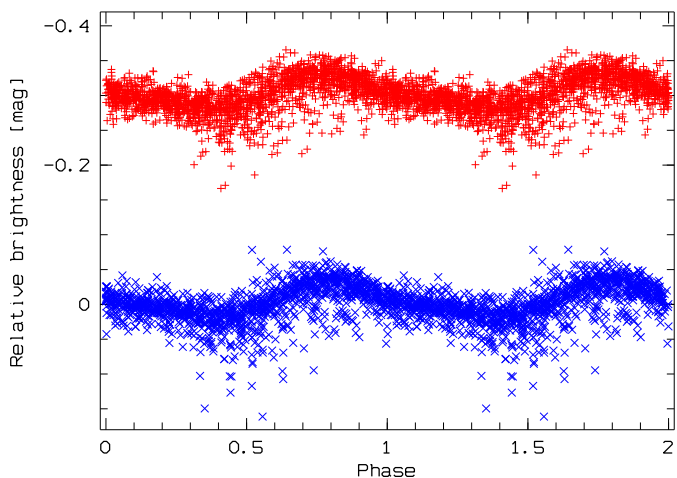}
\caption{
Variability of \object{$\eta$ Cen} with the 1.5562\,c/d frequency (top: 
red passband, bottom: blue passband; phase zeropoint is arbitrary).  
In both light curves, note the deviation from sinusoidality and the 
non-random distribution of `outliers', which are probably related to 
the mass loss process rather then of instrumental origin.  
} 
\label{etaSTEFLlc} 
\end{figure}

Frequencies below about 1 c/d pose a challenge because this range is
home to slow nonperiodic stellar variations including outbursts, beat or
similar phenomena related to higher frequencies, and possible
instrumental drifts.

In summary, frequencies between 1\,c/d and 8\,c/d can be determined
with high confidence.  Analysis, performed using real as well as
synthetic data, in this frequency range of constant stars and very
low-amplitude variables of comparable brightness shows (Baade et al.,
in prep.; see also Pigulski et al., submitted to A\&A) that the above
methods can reliably detect BRITE variables with amplitudes down to
0.5 mmag.  (Here and throughout the rest of the paper, all amplitudes
correspond to the amplitude of a sine function, i.e., are
semi-amplitudes, unless stated otherwise.)

\section{Results} 
\subsection{\object{$\eta$ Cen}} 
\label{etaresults}

The AOV spectrum (Fig.\,\ref{etaAOV20}) is dominated by the {\v S}tefl
variability with 1.5562\,$\pm$\,0.0001\,c/d and its aliases, some of
which are stronger than the AOV line due to the satellites' orbital
frequency. Because of the very large amplitude (see
Fig.\,\ref{etaSTEFLlc}), no other features were detectable with
satisfactory confidence.  Therefore, the 1.56-c/d variability was
removed by prewhitening.

\begin{figure}
\includegraphics[width=6.2cm,angle=-90]{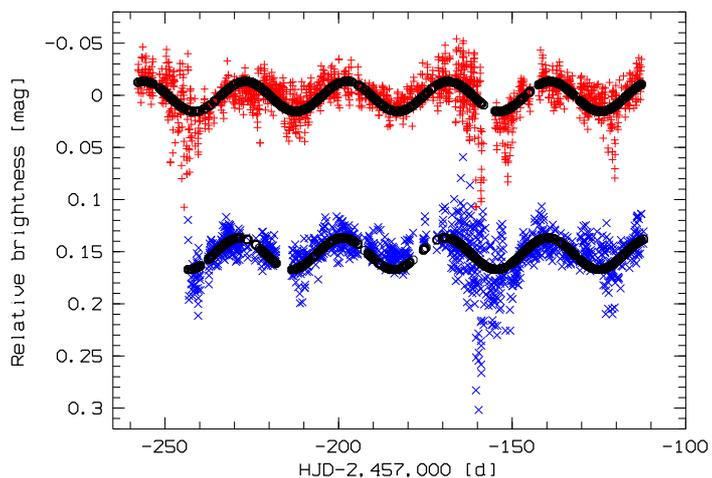}
\caption{
The light curve (red: $+$; blue: $\times$, vertically offset by 
 0.15\,mag) of 
\object{$\eta$ Cen} after prewhitening for the 1.5562\,c/d {\v S}tefl 
frequency (without this prewhitening applied the structure of the 
light curve is extremely similar). Overplotted are fits with the 
frequency, 0.034\,c/d which corresponds to the difference between the 
1.732\,c/d and 1.767\,c/d variations.  Note the 
apparent outburst around day $-160$ (perhaps, a second one happened at 
day $-245$) and the enhanced scatter near extrema and especially minima. 
The phase of the blue variability may be slightly different from the red one.  
In view of the additional ephemeral variations this is not considered 
significantly established. 
}
\label{etaPRW}
\end{figure}

Considering the power of the subtracted variability (blue amplitude:
22.8\,mmag; red amplitude 20.6\,mmag), the remaining peak-to-peak
range was unexpectedly large ($> 200$\,mmag in the blue and $>
150$\,mmag in the red passband). To a good fraction, it is due to another
large-amplitude, but much slower variability with a frequency of
0.034\,c/d (Fig.\,\ref{etaPRW}).

\begin{figure}
\includegraphics[width=6.2cm,angle=-90]{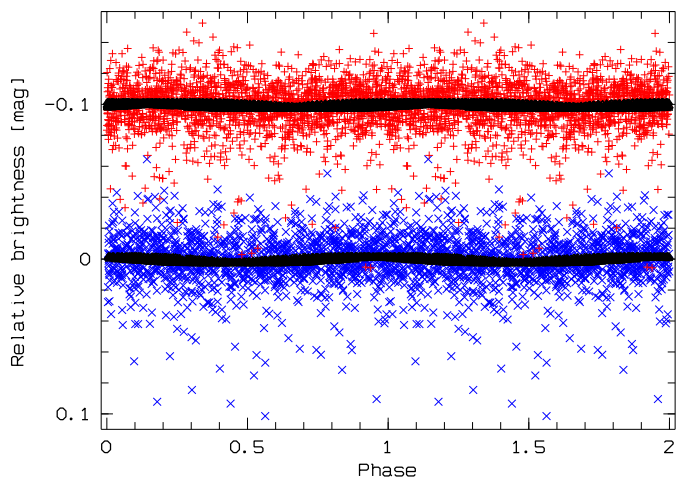}
\caption{
Variability of \object{$\eta$ Cen} with the 1.732-c/d frequency.  
The red-channel 
data are plotted at the top, blue data at the bottom.  The signals 
from the 1.5562- and 0.034-c/d variations have been subtracted.  
A sine curve fitted with the 1.732-c/d frequency is overplotted.  
}
\label{eta17LCa}
\end{figure}

\begin{figure}
\includegraphics[width=6.2cm,angle=-90]{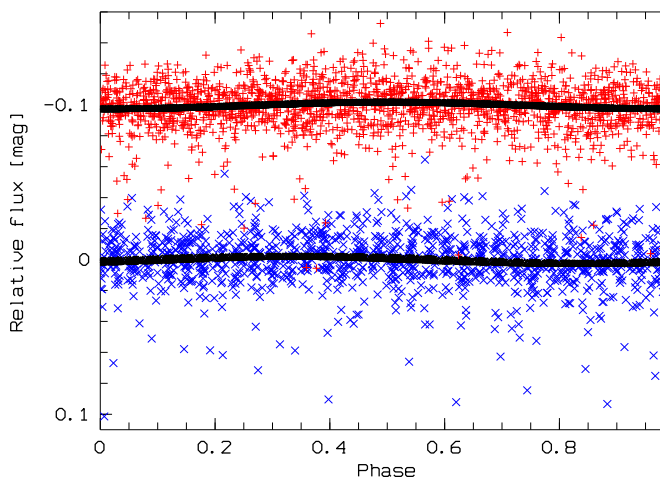}
\caption{
Same as Fig.\,\ref{eta17LCa} except for frequency 1.767\,c/d.
}
\label{eta17LCb}
\end{figure}

\begin{figure}
\includegraphics[width=6.2cm,angle=-90]{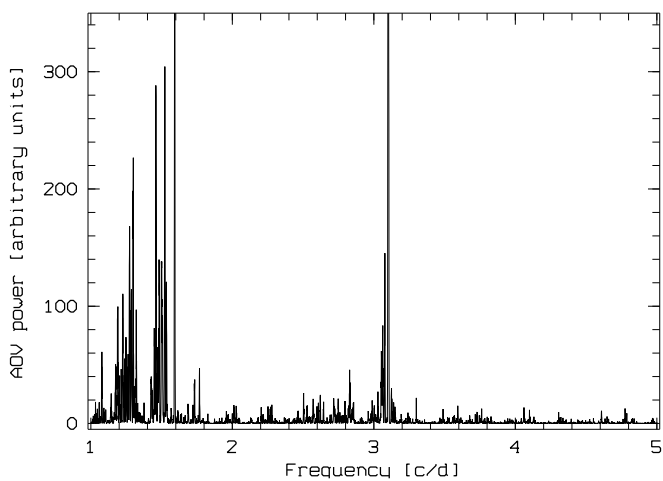}
\caption{
Product of the AOV spectra (after prewhitening for 1.5561 and 0.034\,c/d) 
of the blue and the red passband.  Note the occurrence of frequency groups 
and the relative isolation of the peaks at 1.732 and 1.767\,c/d.
}
\label{etaAOV17}
\end{figure}

This 0.034-c/d signal was also subtracted but the remaining
variability was still large.  AOV period searches found rapid
variations with frequencies 1.732 and 1.767\,c/d. The light curves are
presented in Figs.\,\ref{eta17LCa} and \ref{eta17LCb}.  While the two
corresponding AOV features seem to be relatively isolated
(Fig.\,\ref{etaAOV17}), most of the other frequencies come in groups
(Sect.\,\ref{freqgroups}; Figs.\,\ref{etaAOV17} and \ref{etaAOV8}).
In Table \ref{etafreq} only the frequencies with the largest amplitude
in their respective group are included.

Inspired by the (spectroscopic) case of \object{$\mu$ Cen}, a relation
of the 0.034-c/d variation to higher frequencies was searched for and
quickly found.  To within the errors, 0.034\,c/d is the difference
between the two frequencies of 1.732\,c/d and 1.767\,c/d.  But the
light curve does $\underline{\rm not}$ exhibit a beat pattern (which
in view of the low amplitudes and the presence of larger-amplitude
variations is not, in fact, expected).  Rather, the low-frequency
variability is very well reproduced by a sine curve with a frequency
corresponding to the difference between the two 1.7-c/d
frequencies. With 15\,mmag, the amplitude of the 0.034-c/d variations is
more than five times as large as either of the two higher frequencies
(Table \ref{etafreq}).

Another interesting finding is that the light-curve minima show
somewhat more scatter than the maxima.  Moreover, this scatter seems
to vary from minimum to minimum.  Some of these anomalies have
event-like character (Fig.\,\ref{etaPRW}).  

\begin{figure}
\includegraphics[width=6.7cm,angle=-90]{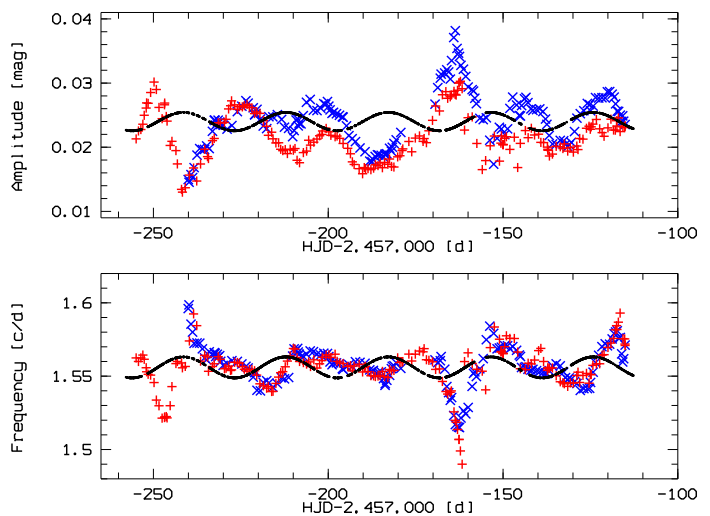}
\caption{
Time dependence of the frequency (lower panel) and amplitude (upper panel) 
of the 1.5562-c/d variability in \object{$\eta$ Cen} (red passband: $+$; 
blue passband: $\times$).  Frequencies and amplitudes were determined as 
sliding averages over 3-day intervals from sine fits.  For comparison, 
the sine curve fitted to the light curve in  Fig.\,\ref{etaPRW} is 
overplotted with arbitrary scaling and vertical offset (note that the 
magnitude scale is inverted).  
}
\label{etaFREQamplSTEFL}
\end{figure}

Because the 1.56-c/d {\v S}tefl frequency belongs to some
extra-photospheric process, its properties were examined in more
detail, and its value and amplitude were derived from sine fits in a
sliding window of 3 days' width. This window covers 4.5 cycles of the
1.5562-c/d variability so that the accuracy is reduced w.r.t.\ the
full 135-150-d range.  The analysis shows that the error margin stated
above for the frequency (0.0001\,c/d) is very misleading and only
valid on average.  As Fig.\,\ref{etaFREQamplSTEFL} (lower panel)
suggests, the temporal variability of the frequency has an amplitude
that is well over an order of magnitude larger: Only when averaged
over many cycles is it roughly periodic.  The amplitude is also
variable and fluctuates between $<$15 and $>$35\,mmag
(Fig.\,\ref{etaFREQamplSTEFL}, upper panel).

The comparison of Fig.\,\ref{etaFREQamplSTEFL} with Fig.\,\ref{etaPRW}
reveals stunning correlations of both frequency and amplitude of the
exo-photospheric {\v S}tefl variation with the star's mean global
brightness.  But there does not seem to be a fixed phase relation, as
can be deduced from the inclusion in Fig.\,\ref{etaFREQamplSTEFL} of
the sine fit to the light curve in Fig.\,\ref{etaPRW}.  Amplitude and
frequency of the {\v S}tefl variability appear crudely anticorrelated
but again with significant phase shifts.  It is important to recall
that the data in Fig.\,\ref{etaFREQamplSTEFL} were derived from
sliding averages over 3 days, i.e.\ about five cycles of the
1.5562\,c/d variability. That is, these figures do not contain
information about variations {\it with} but {\it of} the {\v S}tefl
frequency.

A frequency search over the range 1-7\,c/d identified a few more
frequencies well above the noise floor (see Table \ref{etafreq}).  All
of them belong to frequency groups (Sect.\,\ref{freqgroups}) and are
their respective strongest member.  There are further frequency groups
but their strongest features do not fulfill the adopted significance
criteria.

\begin{table}
\caption{Frequencies in \object{$\eta$ Cen}.  `Blue' and `Red' denote the blue 
and red passband. The errors are about 0.001\,c/d and 0.7\,mmag; `:' denotes 
uncertain detections.}            
\label{etafreq}      
\centering                        
\begin{tabular}{c c c c c}        
\hline\hline              
& \multicolumn{2}{c}{Blue}  &    \multicolumn{2}{c}{Red} \\
ID    &  Frequency     & Amplitude &  Frequency   &  Amplitude  \\   

      &[c/d]           & [mmag]    &  [c/d]       &  [mmag]     \\ 
\hline                       
$f_1$ & 0.0338         & 16.1      & 0.0341       &  14.3       \\
$f_2+$& 1.2242         & 4.6       & 1.2850       &  4.7        \\
$f_3+$& 1.5000         & 7.6       & 1.5230       &  6.3        \\
$f_4+$& 1.5562         & 22.8      & 1.5562       &  20.6       \\
$f_5$ & 1.7314         & 2.4       & 1.7333       &  2.1:       \\
$f_6$ & 1.7672         & 3.9       & 1.7554       &  1.7:       \\
$f_7+$& 2.8184         & 3.9       &              &             \\
$f_8+$& 3.1015         & 6.9       & 3.1076       &  6.8        \\
\hline                                   
\end{tabular}
\tablefoot{A `$+$' in a frequency ID indicates that the frequency 
is the one with the highest amplitude in a group (Sect.\,\ref{freqgroups}, 
after prewhitening for $f_4+$).  As the example of $f_4+$ shows, such 
frequencies may not be constant with time (Sect.\,\ref{etaresults} and 
Paper II).}  
\end{table}

\subsection{\object{$\mu$ Cen}} 

\begin{figure}
\includegraphics[width=6.2cm,angle=-90]{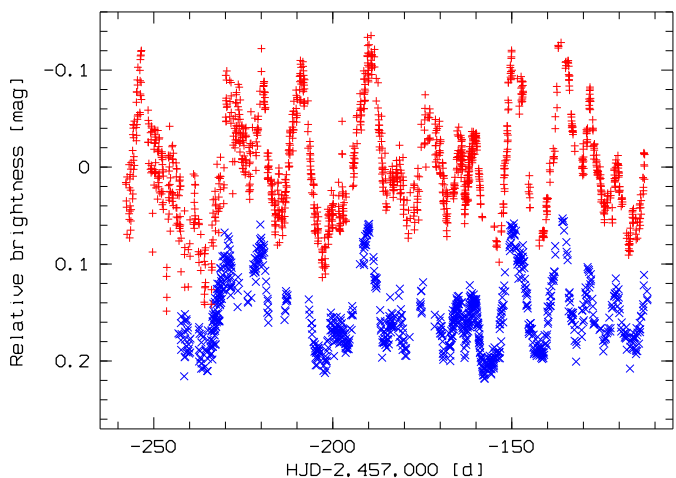}
\caption{
The BRITE light curve of \object{$\mu$ Cen} (in instrumental, 
mean-subtracted magnitudes; red passband: $+$; blue passband: 
$\times$, vertically offset by 0.15\,mag).  
}
\label{muLCbrite}
\end{figure}

\begin{figure}
\includegraphics[width=6.2cm,angle=-90]{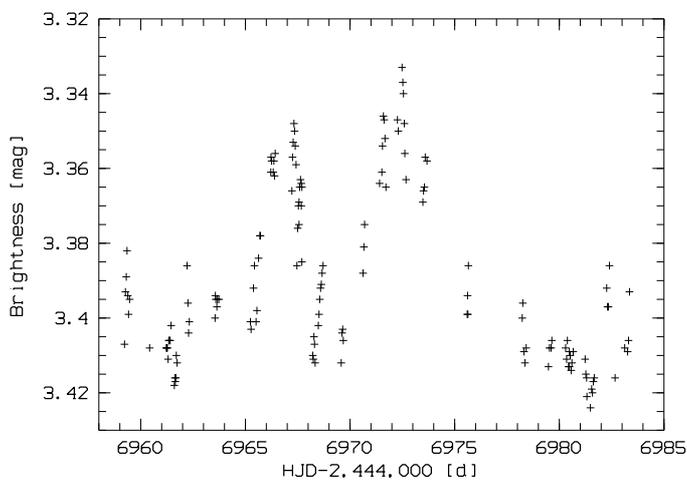}
\caption{
The Str{\" o}mgren {\it b} light curve of \object{$\mu$ Cen} as measured by 
\cite{1989A&AS...81..151C}.  
}
\label{muLCcuypers}
\end{figure}

\begin{figure}
\includegraphics[width=6.2cm,angle=-90]{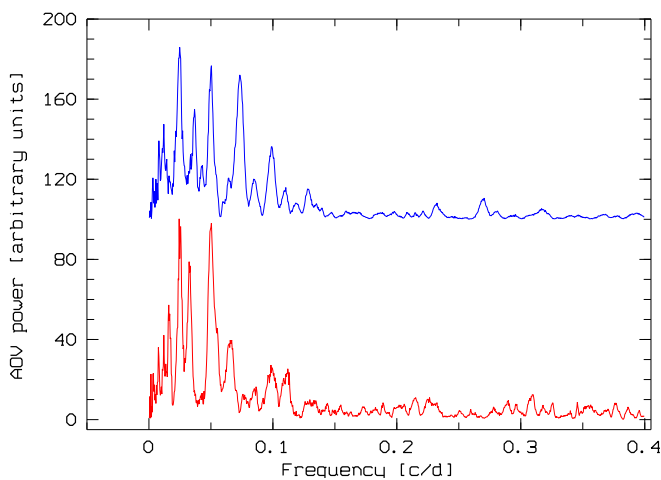}
\caption{
AOV spectra of \object{$\mu$ Cen} (top: blue channel; bottom: red channel).  
}
\label{muAOV2}
\end{figure}

The light curve of \object{$\mu$ Cen} (Fig.\,\ref{muLCbrite}, see also
Fig.\,\ref{muLCcuypers}) shows huge peak-to-peak amplitudes of about
200 and 250 mmag in the blue and the red passband, respectively.  The
time scales are much longer than 1\,d, and in the AOV spectra there
are several prominent features below 0.15\,c/d (Fig.\,\ref{muAOV2}),
which practically prevent a meaningful search for higher frequencies
because elementary confirmation from phase plots is impossible.
Frequencies roughly shared (to within 0.005\,c/d; these features are
fairly broad) by both passbands include 0.025, 0.032, and 0.041\,c/d.
Because there is no such low-frequency power in observations with the
same satellites of \object{$\eta$ Cen}, it should be intrinsic to
\object{$\mu$ Cen}.  Folding the data with the corresponding
hypothetical periods did not yield regularly repeating light curves.
Among these frequencies, only 0.032\,c/d may have a numerical
relationship, or more, to the spectroscopic beat frequencies of 0.018
and 0.034\,c/d found by \cite{1998A&A...336..177R}.  Only 
contemporaneous spectroscopy could provide evidence.

In a next step, the AOV peaks below 0.15\,c/d were recursively removed
by prewhitening. The noise level was reached with less than 10
iterations. The remaining variations in magnitude were still
$\pm$50\,mmag.  This is very much higher than the instrumental noise
and exceeds typical photometric amplitudes of nonradial pulsations in
Be stars by an order of magnitude.  Nevertheless, a detailed AOV
analysis of the prewhitened BRITE data did not identify a single
significant candidate frequency in the range 1-7\,c/d.  This may be
indicative that much of the residual $\pm$50-mmag variability results
from (circum-)stellar noise.

\subsection{Frequency groups}
\label{freqgroups}

The first Be star with identified groups of frequencies was
\object{$\mu$ Cen} \citep{1998A&A...336..177R}. They consist of
coherent (at least those showing beat processes) stellar
eigenfrequencies, four around 2.0 c/d and two near 3.6 c/d.  Limited
sensitivity may well have prevented the detection of more frequencies
in each group. There are no obvious inter-group relations.  In order
to distinguish these \object{$\mu$ Cen}-style frequency groups from
the groups described below they will be called Type I frequency
groups.

\begin{figure}
\includegraphics[angle=-90,width=8.5cm]{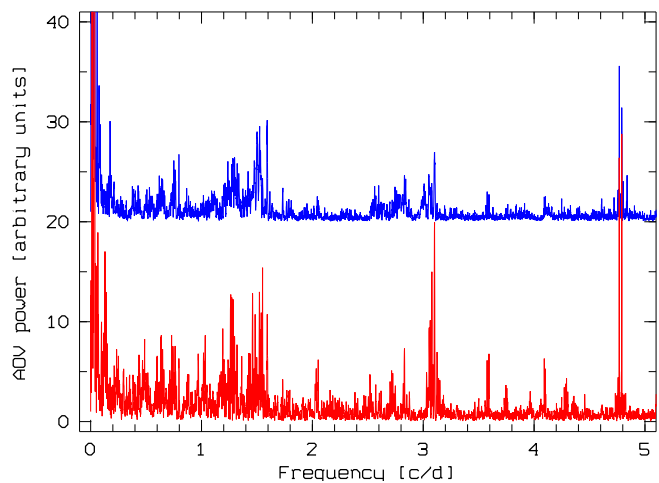}
\caption{
Blue (top) and red (bottom) AOV spectra of \object{$\eta$ Cen} after 
pre-whitening for the 1.5562\,c/d\,{\v S}tefl frequency; see also 
Fig.\,\ref{etaAOV20}.  Note the numerous features (`grass') around 1.55 
and 3.1\,c/d but also below both frequency regions.  The complex close 
to 4.8\,c/d corresponds to one-third 
of the satellites' orbital frequency ($f_{o}/3$).  Compare this figure 
also to Fig.\,\ref{etaAOV20}. 
}
\label{etaAOV8}
\end{figure}

\begin{figure}
\includegraphics[angle=0,width=9.0cm]{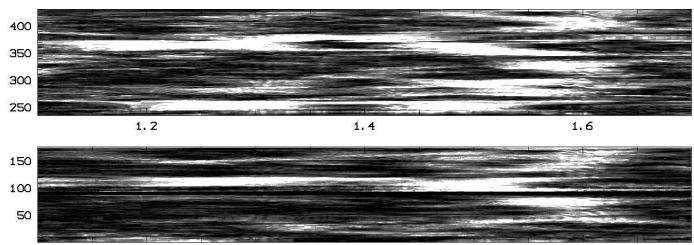}
\caption{
Time-frequency diagram (bottom image: blue passband; top image: red
passband) of the Type II frequency group around 1.55\,c/d in
\object{$\eta$ Cen}, after pre-whitening for the 1.5562\,c/d
frequency.  AOV spectra with a sampling of 0.0005\,c/d were calculated
over 3-d sliding averages.  The frequency interval shown (abscissa) is
1.1-1.7\,c/d.  For both sub-images the AOV power (white is highest)
was separately normalized to the respective mean power to ease
comparison.  The ordinates (in days) have arbitrary zeropoints.  The
two data strings have different starting dates but end at about the
same time (cf.\ Fig.\,\ref{etaPRW}). 
}
\label{stefl2DgrassPRW}
\end{figure}

In the AOV spectrum of \object{$\eta$ Cen}, numerous spikes (`grass')
occur close to the 1.56-c/d {\v S}tefl frequency.  Similar but weaker
features can be found around 3.1 c/d.  They are not direct harmonics
of any individual feature near 1.56 c/d so that one might rather speak
of `group harmonics'.  Figs.\,\ref{etaAOV17} and \ref{etaAOV8} give an
overview of these Type II frequency groups in \object{$\eta$ Cen}.
They also show further tufts of grass, mainly shortward of 1.55 and
3.1\,c/d (see Table \ref{etafreq} for more examples).  Because it is
based on data strings of no more than 3 days,
Fig.\,\ref{stefl2DgrassPRW} cannot resolve the Type II frequency
groups around 1.2 and 1.55\,c/d in \object{$\eta$ Cen} into discrete
AOV features.  But there are major redistributions with time of the
total power.  The high structural similarity of the blue- and the
red-passband diagrams in Fig.\,\ref{stefl2DgrassPRW} confirms in a
more encompassing way that this noise is intrinsic to $\eta$ Cen and
not instrumental.

A probably related result is that, although many of the AOV spikes of 
frequency-group members are stronger than those of the 1.7-c/d frequencies, 
they do not pass the simple empirical significance check described in 
Sect.\,\ref{method}.

Several mechanisms are conceivable that could lead to frequency
patterns that appear similar to this description of Type II frequency
groups but actually are spurious.  They were tested and eliminated one
by one: \begin{list}{$\bullet$}{\itemsep=0mm\topsep=0mm\partopsep=0mm}

\item 

There could be instrumental noise.  But: Simultaneous blue and red
data obtained with different satellites show the same general
phenomenon while the cross-color match of individual features is poor.
At the same time, the AOV spectra of \object{$\mu$ Cen} of data extracted from
the same CCD images do not show similar features at the frequencies in
question.

\item

Prewhitening with an incorrect (not genuine) frequency can introduce
noise, the 1.556-c/d frequency is in the range concerned, and it is
variable so that subtraction of a constant frequency is necessarily
imperfect.  But: The features are already visible before prewhitening,
and using different trial frequencies did not alter the nature of the
frequency groups.

\item

Deviations from sinusoidality can introduce artifacts during
prewhitening.  But: The prewhitening procedure used does not make any
assumptions about the shape of the variation. 

\item

Phase jumps due to outbursts can cause single frequencies appear
multiple.  But: The other variabilities show no evidence of such
jumps.

\end{list} Moreover, for each of the above concerns to become reality,
the frequency ranges seem far too broad.

Because the AOV spectra of \object{$\mu$ Cen} are more `populous', it
is not obvious whether or not Type II frequency groups exist in this
star, too.  The best candidate appears around 3.4\,c/d.

\section {Analysis of the results} 

Be stars differ from other pulsating early-type stars in that, as
pulsators, they are players but in addition also produce their own
stages in the form of the circumstellar disks, which may echo the
stars' activity.  This adds pitfalls to the analysis but also offers
additional diagnostics.  It is vital to not lose track of this
duality.  Extensive model calculations by \cite{2012ApJ...756..156H}
provide a very useful illumination of the photometric response to
varying amounts of circumstellar matter as seen at different
inclination angles.

\subsection{\object{$\eta$ Cen}} 

The photometric variability of \object{$\eta$ Cen} contains 
major circumstellar components:
\begin{list}{$\bullet$}{\partopsep=0mm\topsep=0mm\itemsep=0mm} 
\item

The $\pm$3\% variation of the frequency with the highest amplitude,
1.56\,c/d (Fig.\,\ref{etaFREQamplSTEFL}, lower panel), is not
plausibly reconcilable with stellar pulsation although the mean value
falls into the range of plausible $g$-modes.  Since the frequency
itself is slightly lower than the frequency of the strongest stellar
pulsations, the classification as circumstellar {\v S}tefl frequency
is confirmed.
 
\item 

The 0.034-c/d phased light curve shows noise beyond photon statistics and
with subtle asymmetry: During minima it is slightly larger than
during maxima.  Because \object{$\eta$ Cen} exhibits weak shell-star
signatures \citep{2006A&A...459..137R} and is observed nearly through
the plane of the disk, this probably means that matter is elevated
above the photosphere and injected into the disk, where it removes
light from the line of sight. This is shown in Fig.\,\ref{etaPRW}.  In
other stars, sudden increases in photometric noise have been observed
to precede major brightenings \citep[e.g.,][]{2009A&A...506...95H}.

\item 

Figure \ref{etaPRW} also suggests that there was a discrete event in
both passbands around day $-160$ and possibly a second one near day
$-245$.

\item 

Apart from outbursts, regularly varying amounts of near-stellar matter
in the disk could reprocess the stellar light and make part of it
detectable along the line of sight.  This would imply that the
0.034-c/d variation drives a quasi-permanent base mass loss.

\item

Similar to the 1.56-c/d peak, many of the other stronger peaks in the
AOV spectrum are surrounded by numerous additional features forming a
group (Sect.\,\ref{freqgroups}; Figs.\,\ref{etaAOV17} and
\ref{etaAOV8}).

\end{list}

\noindent Based on the photometry alone, the two frequencies near
1.7\,c/d are only marginal detections.  But the presence of the same
frequencies in the {\sc Heros} spectroscopy, which identified them as
NRP modes (Sect.\,\ref{etaspec}), clearly boosts them to significance.
The agreement of the difference between these two frequencies and the
0.034-c/d variation suggests that all three are linked.  If so
(Sect.\,\ref{nature034} resumes and completes this discussion), the
regularity of the mean 0.034-c/d light curve indicates that the
1.7-c/d frequencies have a better phase coherence than the 1.5562-c/d
{\v S}tefl frequency.  The slow variability shown in
Fig.\,\ref{etaPRW} extends over 250 cycles of the 1.7-c/d variations.
Frequency noise at the 3\% level of the 1.7-c/d pulsations would not
let the 0.034\,c/d light curve repeat as well as it actually does.

It is interesting to note that the amplitudes of the two 1.7-c/d
variations are only 2-3\,mmag while the slow 0.034\,c/d variation
reaches nearly 15\,mmag, i.e., about three times as much as the
amplitude sum of the two fast variations (see Table \ref{etafreq}).

If the 0.034-c/d variation does result from the interaction of two NRP
modes, they would become the root cause also of the remainder of this
complicated variability pattern: The individual cycles of the
0.034-c/d light curve (Fig.\,\ref{etaPRW}) and even the residuals
from the mean amplitude are very well matched by the variability of
the {\v S}tefl frequency and its amplitude.  This suggests a fairly
tight coupling between this pulsation-related process and the
circumstellar {\v S}tefl process.  But the large phase wiggles require
the coupling to be quite elastic.

\subsection{\object{$\mu$ Cen}} 

A particularly strong illustration of the need to take into account
circumstellar variability is provided by this star.  The combination
of long timescale and very large amplitude of the dominant photometric
variability makes the disk a much better candidate for the origin of
this variation than the star.  The variability is not due to a
variable flux from the star.  Rather, the varying amount of recently
ejected matter in the near-stellar part of the disk reprocesses the
presumably just very slightly intrinsically variable stellar flux and
redirects part of it to the observer's line of sight.  Because the
inclination angle is low \citep[$\sim$20 deg;
][]{2001A&A...369.1058R}, the effect is very pronounced
\citep{2012ApJ...756..156H} and the amount of extra light provides a
crude measure of the amount of extra near-stellar matter.

If \object{$\mu$ Cen} possesses any short-period low-amplitude
variability, it would be difficult to detect against the flash lights
of the large circumstellar activity.  This parasitic light may be
powered off by the temporary dispersal of the disk \citep[cf.\
][]{1986ApJ...301L..61P, 1988A&A...198..211B}.  However, what then
becomes visible may not be representative of the active phases.

\subsection{Frequency groups} 

By definition, Type I frequency groups consist of stellar NRP modes
only, and \object{$\mu$ Cen} is the first member of this class.  It is
to be seen whether this class will remain so elusive.  This
photometric study did not identify any Type I group in either star
(the two 1.7-c/d frequencies in $\eta$ Cen do not seem to qualify as a
group).

Because of their proximity in \object{$\eta$ Cen} to the
demonstratedly circumstellar {\v S}tefl frequency, Type II frequency
groups, too, may be rooted in the near-photospheric environment. The
drifting power in Fig.\,\ref{stefl2DgrassPRW} reinforces this
conclusion quite strongly (Sect.\,\ref{nature034}).  In fact,
circumstellar processes are not expected to be strictly periodic, they
must be intrinsically noisy, and there will be nonlinearities that
lead to higher-frequency components.  This is a good description of
the `grass' close to the {\v Stefl} frequencies as well as of the
group harmonics seen in BRITE AOV spectra of \object{$\eta$ Cen}.
Paper II elaborates on the appearance and nature of Type II frequency
groups in much additional detail.

Neither the definitions of Types I and II nor the proposed explanation
as stellar and circumstellar, respectively, exclude the possibility
that a star has frequency groups of both types.

\subsection{Comparison of BRITE and earlier ground-based
photometry} 

The following subsections only discuss selected earlier
observations. For both stars, a more complete overview and critical
evaluation is provided by \cite{2003A&A...411..229R}.

\subsubsection{\object{$\eta$ Cen}} 

Because of its very large amplitude, the 1.56-c/d variation was
detected by everyone with adequate data \cite[e.g.,
][]{1989A&AS...81..151C, 1995A&A...294..135S}.  However, the phase
instability sometimes made subharmonics look like a better match to
the observations, causing some observers to conclude that this
variability of \object{$\eta$ Cen} is due to some co-rotating
structures.  The 0.52-c/d variant in particular was refuted by
\cite{2003A&A...411..229R}, which is confirmed by the BRITE
observations.

The amplitude of the 0.034-c/d light curve is only slightly lower
(15\,mmag vs.\ 20\,mmag).  But the timescale of a month means that 
ground-based observations do not stand a chance of detecting it.  

\subsubsection{\object{$\mu$ Cen}} 
\label{etaOLDspec}
\cite{1989A&AS...81..151C} published the only major photometric
dataset of \object{$\mu$ Cen} obtained from the ground (Fig.\,\ref{muLCcuypers}).
The light curve with frequency 0.476 c/d derived by those authors had
a rather extreme, ragged shape, and is not at all supported by the
BRITE observations.

This work and that of \cite{1989A&AS...81..151C} agree that the
photometric variability does not contain any significant trace of the
periods found in the long series of {\sc Heros} spectra
\citep{1998A&A...336..177R}.  The large signal can rather be thought
of as some circumstellar light pollution.  It is, therefore, kind of a
curious footnote that the photometry by \cite{1989A&AS...81..151C} --
along with combined low- and high-order line-profile variability
\citep{1984A&A...135..101B} -- provided the incentive to put
\object{$\mu$ Cen} on the {\sc Heros} target list, which resulted in
the discovery of the most complex spectroscopic pulsation pattern
found to date in any Be star \citep{1998A&A...336..177R}.

The comparison of Figs.\,\ref{muLCbrite} and \ref{muLCcuypers}
suggests that over as much as 27 years (about 20,000 half-day
pulsation cycles), the overall character of the photometric
variability of \object{$\mu$ Cen} has not changed.

\subsection{Comparison of BRITE photometry with earlier
spectroscopy} 

\subsubsection{\object{$\eta$ Cen}}
\label{etaspec}

\cite{2003A&A...411..229R} detected frequencies of 1.56 c/d, 1.73 c/d,
and 1.77 c/d.  The match with BRITE is excellent - even after
$\sim$2 decades.  \cite{2003A&A...411..229R} identified the
possibility that beating of the two most rapid variations could
lead to outbursts.  But the series of spectra available to them was
not sufficient to conclude this with acceptable confidence.  

BRITE photometry alone does not provide any direct insights into the
nature of the pulsations and, therefore, depends on spectroscopic
support.  \cite{2003A&A...411..229R} found the line-profile
variability associated with 1.73\,c/d to be of the same
type as in almost all other bright Be stars and attributed it to
nonradial $\ell = m = +2$ $g$-modes. In a later re-analysis of the 
{\sc Heros} spectra, Zaech \& Rivinius (unpublished) confirmed 
that both 1.7-c/d variations belong to this type.  

The MiMeS project studied 85 Be stars \citep{2014arXiv1411.6165W}.  In
none of them was a large-scale magnetic field found. \object{$\eta$
Cen} was not included in the sample but it was observed by ESPaDOnS
(1$\times$) and HARPSpol (9$\times$). After careful reduction,
the 9 measurements with HARPSpol \citep{2011Msngr.143....7P} on 30
April 2014 were coadded and mean Least-Squares Deconvolved \citep[LSD;
][]{1997MNRAS.291..658D} profiles were extracted for the unpolarized
(Stokes $I$), the circular polarized (Stokes $V$) and diagnostic null
profiles. The longitudinal magnetic field measured from the LSD
profile \citep[$9\pm17$\,G; ][]{2014arXiv1411.6165W} provided no
evidence of a magnetic detection. However, a marginal detection was
established based on $\chi^2$ statistics (Donati et al. 1992) due to a
series of pixels between about $-$50 and 200\,km/s (the unpolarized
LSD profile spans a velocity range between $-$400 km/s to 330 km/s)
that have signal outside of the error bars. This test computes the
false alarm probability (FAP) that measures the probability that the
observed $V$ signal inside the line profile differs from a null
signal. In this case, there is sufficient signal in this line such
that the FAP is low enough to consider this a marginal detection
(10$^{-6}$ $<$ FAP $<$ 10$^{-4}$). That is, the hypothesis that the
line profile can be explained entirely by noise is in poor agreement
with the observation. Since no excess or similar signal is found in
the null profile, the feature seen should not be instrumental.

Only additional observations can remove the attribute `marginal
detection'.  It is hoped that the ongoing BRITE spectropolarimetric
survey \citep{2014sf2a.conf..505N} can achieve this among the
$\sim$600 targets to be observed brighter than V=4\,mag.  The survey
data taking is expected to finish in the first half of 2016.

\subsubsection{\object{$\mu$ Cen}} 

The analysis methods described above did not return a single one of
the six spectroscopic frequencies (four of them near 2 c/d and two
near 3.6 c/d) found by \cite{1998A&A...336..177R} and classified as
$\ell=m=+2$ $g$-modes.  Therefore, a special `targeted' search for
these frequencies was performed in the BRITE data. Weak AOV features
occur at 1.991 and 2.027\,c/d.  The three strongest spectroscopic
variations found by Rivinius et al.\,have frequencies of 1.988, 1.970,
and 2.022\,c/d. That is, at least the second strongest spectroscopic
variation has no BRITE counter part.

\object{$\mu$ Cen} was observed with ESPaDOns on 24 February 2010 as
part of the MiMeS project \citep{2014arXiv1411.6165W}.  A new
reduction of the data confirmed the non-detection at a level of
B$_{\rm z}$=9$\pm$12\,G, which is also consistent with the
results from the $\chi^2$ statistics.

Finally for these comparisons, not even a targeted search could detect
the {\v S}tefl frequency at 1.61 c/d \citep{2003A&A...411..229R} in
the red passband.  In the blue AOV spectrum, a spike at 1.615\,c/d is
nearly the highest peak in what might be a Type II frequency group (or
a high-frequency artifact caused by the large-amplitude slow
variations).

\section{Discussion}
\subsection{Synopsis of the variabilities of $\eta$ and \object{$\mu$ Cen} }

\begin{table}
\caption{Comparison of $\eta$ and \object{$\mu$ Cen}.  Main sources: 
\cite{1998A&A...333..125R}, \cite{2001A&A...369.1058R}, 
\cite{2003A&A...411..229R}, this study, and {\it SIMBAD}.  N/A = not available}
\label{COMPAREmueta}     
\centering           
\begin{tabular}{l | l | l}        
\hline\hline                 
Property                        & \object{$\eta$ Cen} & \object{$\mu$ Cen} \\    
\hline                        
MK type                         & B2 Ve              & B2 Vnpe \\      
$v$ sin $i$ [km\,s$^{-1}$]      & 300                & 155   \\
Inclination angle [degree]      & 85                 & 19 \\
Rotation frequency [c/d]        & 1.7                & 2.1 \\
Low (beat?) frequencies [c/d]   & 0.034              & 0.018; 0.034  \\
Outbursts synchronized          &                    &  \\
with slow light curve?          & Y                  & Y   \\
Frequencies of                  & 1.732; 1.764       & 1.970; 1.988; \\
interacting NRP modes [c/d]     &                    & 2.022 \\ 
Degree and order of             & 2,$+2$; 2,$+2$     & 2,$+2$; 2,$+2$; \\
interacting NRP modes           &                    & 2,$+2$ \\
Semi-amplitudes of              &                    & \\
interacting modes [mmag]        & 3; 2.5             & N/A \\
Peak-to-peak amplitude          &                    &  \\
of outbursts [mmag]             & 100                & 250 \\
Main circumstellar {\v S}tefl   &                    & \\
frequency [c/d]                 & 1.56               & 1.6 \\
{\v S}tefl frequency detected   &                    &   \\
by BRITE?                       & Y                  & N \\
{\v S}tefl frequency permanent? & Y                  & N \\
\hline                                   
\end{tabular}
\tablefoot{The identification of also the second 1.7-c/d variability 
in the {\sc Heros} spectra of \object{$\eta$ Cen} as $\ell = m = +2$ is by 
Zaech \& Rivinius (unpublished).}
\end{table}

Table \ref{COMPAREmueta} juxtaposes a number of properties of $\eta$ and $\mu$
Cen.  At a first glance, the spectroscopic and photometric
appearances of their pulsations seem confusing, if not contradictory.
However, consideration of the difference in inclination angle reveals
simple and well-known systematics: 
\begin{description}{\topsep=0mm}
\item[\object{$\eta$ Cen}] shows weak shell-star properties.  
This equator-on perspective 
favors the photometric detection of quadrupole modes, and
freshly ejected matter re-directs relatively little light into the line of
sight.  
\item[$\mu$
Cen] is viewed at a small inclination angle, which facilitates the detection of
the latitudinal component of the horizontal $g$-mode velocity field
\citep{2003A&A...411..229R}. In photometry, much of the equatorially
concentrated light variation (quadrupole modes) is lost to azimuthal
averaging.  Instead, the processing of radiation by regularly newly
ejected matter close to the star leads to a huge photometric signal
that outshines everything else.  
\end{description}

\noindent Both stars suffer enhanced mass loss that is driven by a
slow process assembled from two much more rapid nonradial pulsations.
Because of the nearly edge-on perspective of \object{$\eta$ Cen}, the
variability is manifested in small fadings of this star, in contrast
to \object{$\mu$ Cen}.

A {\v S}tefl frequency is present in the series of {\sc Heros} spectra
of both stars.  But BRITE only found that in \object{$\eta$ Cen},
at a considerable amplitude.  This particular difference will be
addressed in Sect.\,\ref{engine}.  The other difference lies in the
degree of their persistence.  In \object{$\eta$ Cen} the {\v S}tefl
frequency is quasi-permanent while in $\mu$ Cen its presence is
coupled to enhanced line emission and the phase of the pulsational
beat process.  This could mean that the mass loss process in
\object{$\eta$ Cen} is always active but is occasionally enhanced at
extrema of the slow 0.034-c/d variation.

\subsection{Comparison to {\it Kepler} and {\it CoRoT} observations  
\hspace{2cm} 
of Be stars} 

It is of obvious interest to compare the BRITE observations of $\eta$
and \object{$\mu$ Cen} to photometry of Be stars obtained with other
satellites.  For this purpose, the description by
\cite{2015MNRAS.450.3015K} of {\it Kepler} observations of the
late-type (B8) B and probable Be star \object{KIC\,11971405} and the
behavior described by \cite{2009A&A...506...95H} of \object{HD\,49330}
(B0.5 IVe) during an outburst as seen by {\it CoRoT} are considered.

The power spectra of both stars exhibit frequency groups.
\cite{2011MNRAS.413.2403B} were among the first to ask whether such
structures arise from rapid rotation and might even be characteristic
of Be stars.  \cite{2015MNRAS.450.3015K} conclude that frequency
groups in Be, SPB, and $\gamma$ Dor stars can be explained by simple
linear combinations of relatively few $g$-mode frequencies.  For the
explanation of outbursts of Be stars, they adopt the same qualitative
scheme as first developed by \cite{1998cvsw.conf..207R}, namely NRP
beating, and significantly refined by \cite{2014arXiv1412.8511K}.

The appearance of the frequency groups is different during quiescence
and outbursts.  Both \cite{2009A&A...506...95H} and
\cite{2015MNRAS.450.3015K} suggest that during outbursts of Be stars a
large number of pulsation modes flare up and drive the mass loss
\citep[see also][ for \object{HD\,163868}]{2005ApJ...635L..77W}.  Because no
physical explanation is given as to what would trigger such avalanches
of pulsations, it is worthwhile searching for alternate descriptions
of this behavior.

In fact, the BRITE photometry of $\eta$ and \object{$\mu$ Cen} does
offer an alternate view of the frequency groups seen by {\it CoRoT}
and {\it Kepler} \citep[and by {\it MOST} -- ][]{2005ApJ...635L..77W}
if they, too, belong to Type II.  Because of the natural circumstellar
noise, such an interpretation would require much less explanatory
extrapolation into the unknown.  A possibly important commonality is
that the frequency groups in \object{HD\,49330} ($\sim$3\,c/d and
$\sim$1.5\,c/d) and \object{KIC\,11971405} ($\sim$4\,c/d and
$\sim2$\,c/d) as well as \object{HD\,163868} ($\sim$3.4\,c/d and
$\sim$1.7\,c/d) are crudely consistent with the description of 2:1
group harmonics.  {\it CoRoT} and {\it Kepler} were more sensitive
than BRITE is.  Therefore, they do not see only a few "blades" of grass
in their power spectra but may observe a whole "prairie" spearing during
outbursts of Be stars (see Paper II).

\subsection{Modulations with azimuth}

Because the stellar rotation frequencies are of the order of 70\% or
more of the Keplerian frequencies (Sect.\,\ref{relrot}) while the
errors of both are well above 20\%, the values of these frequencies
are not suitable to determine the location of the {\v S}tefl process
beyond confirming it to lie in the star-to-disk transition region.

Table \ref{freqtab} suggests that {\v S}tefl frequencies occur at about
one-half of the maximal Kepler frequency.  At the same time, the
associated velocities are above equatorial velocities, which typically
reach or exceed 70\% of the critical velocity.  This apparent (weak) 
contradiction may be resolved if the azimuthal period of the
structures is a fraction of 360\,degrees.  If the fraction is an
integer, $j$, as in nonradial pulsations, $j = 2$ is a good
guess. Because {\v S}tefl variations are not strictly periodic, $j$
does not have to be an integer.  But it simplifies the discussion to
make such an initial assumption.  In \object{$\eta$ Cen}, the {\v S}tefl
frequency is closely tied to the 0.034-c/d process, which involves a
stellar quadrupole mode.  Therefore, $j = m = 2$ is also on this
ground a good starting assumption.

\subsection{The nature of the 0.034-c/d variation in \object{$\eta$ Cen}}
\label{nature034}

Because of its sinusoidal shape, the 0.034-c/d variability of
\object{$\eta$ Cen} is the most unexpected discovery made by the
observations with BRITE.  Owing to its relation to mass loss, it is
probably the signature of $\eta$ Cen's inner (mass-loss) engine.  But
the design of that engine remains enigmatic, and it may still be
without precedence in Be stars.  Therefore, the following sections aim
at constraining its nature by exploring four very different hypotheses
for its explanation.  
\vspace{1mm}

\noindent
{\it First hypothesis: orbital variability in a binary system}

In a binary system, a sinusoidal light curve can result from the 
distortion of two nearly identical stars.  In this case, the orbital 
period would be two times 29.4\,d, and the separation
of two 9-$M_{\odot}$ stars would amount to about 165 $R_{\odot}$.  This 
large separation would invalidate the hypothesis of tidal distortion.  
For a cool and large companion, 29.4\,d could be the orbital period.  But 
the separation would be too large for reflected light from the 
primary B star to have the observed amplitude.  The absence of any 
extended constant parts of the light curve basically rules out 
partial eclipses.  
\vspace{1mm}

\noindent
{\it Second hypothesis: companion star-induced global disk
oscillations}

In binaries with a separation of order 0.5\,au or more, eclipses may
well not happen.  However, if a companion is not too close to prevent
the formation of a sizeable disk with Be star-typical line emission,
it may induce global disk oscillations.  The associated density waves
carry a significant photometric signal (Panoglou et al., submitted to
MNRAS), which in disks viewed pole-on can reach 100 mmag for strongly
ellipical orbits.  In \object{$\eta$ Cen}, the sinusoidal variation
would probably imply a fairly circular orbit and a very regular
oscillation pattern. In some (but not all) Be binaries, V/R
variations of emission lines track the orbital phase
\citep{2007ASPC..361..274S}.  Because of slow large-amplitude
variations in the strength and overall structure of the emission
lines, the {\sc Heros} spectra could not be used to search for
periodoic V/R variability.  However, this softer version of the
binary hypothesis still faces the objection that matching
radial-velocity variations were not discovered by
\cite{2003A&A...411..229R} and \cite{2006A&A...459..137R}.  A
low-mass sdO companion (see \citet{2012ASPC..464...75R} and
\citet{2012A&A...545A.121K} for the latest candidates put on the still
very short list) might go unnoticed in radial-velocity data but would
often reveal itself through He\,II 468.6 line mission from the region
of the disk closest to it.  The {\sc Heros} spectra do not show this
feature.

\vspace{1mm}

\noindent
{\it Third hypothesis: coupled nonradial pulsation modes}

There is no known mechanism that would work with a single frequency of
0.034\,c/d in the atmospheres of single early-type stars.  However,
the identity, to within the errors, of 0.034\,c/d to the difference
between the two short frequencies at 1.732 and 1.764\,c/d is
reminiscent of what in other stars is called a combination frequency.
In the case of $\eta$ Cen, it would be the difference between two
pulsation frequencies.  But frequency sums occur as well.  The
abundance of sums and differences is differently biased in different
stars but the reason is not known.  \cite{2001MNRAS.323..248W},
\cite{2012MNRAS.422.1092B}, and \cite{2015MNRAS.450.3015K} discuss the
amplitudes of combination frequencies in white dwarfs, $\delta$ Scuti
stars, and $\gamma$ Dor / SPB / Be stars, respectively.  Kurtz et
al.\,conclude that non-linear mode coupling can give combination
frequencies of $g$-modes a higher photometric amplitude than the
parent frequencies.  In their analysis, this results largely from the
coupling of high-order modes to form low-order variations, which
suffer less cancellation across the stellar disk.  But in
\object{$\eta$ Cen} the assumed parent modes are quadrupole modes, and
the amplification factor would be huge with three times the sum of the
two 1.7-c/d variations.  To include these facts would probably require
a siginificant extension of the notion of combination frequencies.
\vspace{1mm}

\noindent
{\it Fourth hypothesis: circumstellar activity}

It is at least an odd coincidence that not only the two stellar
1.7-c/d frequencies differ by 0.034\,c/d but the circumstellar
1.5661-c/d frequency also differs by this amount from its two
strongest neighbouring peaks in the AOV spectra (Table
\ref{etafreq}).  As Fig.\,\ref{etaFREQamplSTEFL} has
illustrated, the 1.5562-c/d frequency is not constant but varies by
more than 0.034\,c/d.  On this basis, one would dismiss the agreement
as an oddity, especially since the entire frequency group, too, of
which 1.5562\,c/d is the strongest member by far, does not seem to be
phase coherent, see Fig.\,\ref{stefl2DgrassPRW}.  However, if the
three frequencies near 1.56\,c/d shifted around in the same fashion,
this argument could be invalid.  In fact, the said two companion peaks
to 1.5562\,c/d are variable in position.  Although the measuring
errors are much larger than for the central peak, their variability is
similar.  But the amplitudes of the frequency variations are nearly
twice as large as the one of the 1.5562-c/d frequency.  This appears
significant since the two substantially weaker stellar 1.7-c/d
variations show much less scatter, which is consistent with the
hypothesis of constant frequencies.

\vspace{1mm}

\noindent
{\it Conclusion}

The ecplising-binary hypothesis can be safely excluded.  The first and
the second one are weakened by the lack of radial-velocity variations.
These first two and the fourth hypothesis have in common that they do
not offer a connection between the 0.034-c/d variability and the
equally large difference in frequency between the two 1.7-c/d stellar
pulsations.  If such an explanation is required, a coupling of the two
pulsation modes is the only useful ansatz among the four options
considered although it may be of a rather different nature than
combination frequencies in other $g$-mode pulsators.

The third hypothesis is the only one that seeks the explanation within
the star rather than in the disk.  Therefore, a strong argument in
support of it derives from the association of the
0.034-c/d frequency with both mass loss and the variability of the
circumstellar {\v S}tefl frequency.  The cause of the mass loss must
be in the photosphere or below it, and the strong response of the {\v
S}tefl frequency to small outbursts that are synchronized with the
0.034-c/d variability is a manifestation of this causal connection.

For these reasons the discussion below only considers the hypothesis of
coupled NRP modes even though it acknowledges that this choice for the
explanation of the inner engine is not unequivocally forced by the
available data.

\subsection{Two mass-loss engines working in series}  
\label{engine}

Since the discovery of {\v S}tefl frequencies, the evidence has been
strong that they are closely related to the mass loss from Be stars
and trace super-photospheric processes.  The combination of {\sc
Heros} spectroscopy and BRITE photometry has established this firmly.
Before BRITE there was some ambiguity as to whether the pulsation
enables the {\v S}tefl frequency or whether the {\v S}tefl frequency
is somehow the rhythm of the engine injecting mass into the
circumstellar disk.  The correlations of the variations of the
frequency and the amplitude of the {\v S}tefl frequency
(Fig.\,\ref{etaFREQamplSTEFL}) with the probably pulsation-related
0.034-c/d variation found by BRITE in \object{$\eta$ Cen} give
convincing support to the former because otherwise the circumstellar
{\v S}tefl process would have to be driving the stellar pulsation.

Moreover, the said ambiguity is not a conflict, and the available
spectroscopy and photometry are not contradicted by the working
hypothesis that both are true: The pulsations enable the {\v S}tefl
frequency, which in a second stage of the mass-loss process feeds the
disk.  This two-stroke process consisting of an inner (pulsations) and
an outer ({\v S}tefl process) mechanism that operates the
mass-transfer from star to disk in \object{$\eta$ Cen} is one
particular realisation of the general two-engine concept.

\subsubsection{The inner engine: Interacting nonradial pulsation modes}
\label{NRPinteract}
\object{$\eta$ Cen} was selected as a BRITE target because of indications of
quasi-permanent mass loss and beating of two NRP modes 
\citep{2003A&A...411..229R} so that any
physical connection between them could be identified and
characterized.  The actual chain found by BRITE is more complex
than anticipated:
\begin{list}{$\bullet$}{\itemsep=0mm\topsep=0mm}
\item
Two nonradial pulsation modes (1.7\,c/d)
\item
combine in an unknown fashion to a much slower variability (0.034\,c/d) 
\item
which drives the mass loss (possibly dissipating more energy than the 
linear sum of the two modes)
\item
and connects to the circumstellar {\v S}tefl process (1.56\,c/d) that 
probably is the mechanism by which the matter organizes itself into 
circumstellar structures. 
\end{list}

\noindent 
In \object{$\mu$ Cen}, no specific mode interaction 
has been inferred other than that outbursts repeat with the difference
frequencies of several NRP modes.  But there
is no direct evidence of a beat phenomenon with its characteristic
envelope describing the variation of the combined amplitude.  It is
not excluded that \object{$\mu$ Cen} is an \object{$\eta$ Cen} analog
but more complicated because of the involvement of at least three
frequencies.

Examples of Be stars with light curves exhibiting a classical beat
phenomenon can be seen in Fig.\,16 of \cite{2007A&A...472..577M} and
in Fig.\,5 of \cite{2008A&A...480..179D}. But the authors do
not report outbursts occuring with the beat frequencies of the four
stars concerned, which are members of the open cluster NGC 330 in the
Small Magellanic Cloud.

At the same time, the very bright and nearby examples \object{28
$\omega$ CMa} and \object{Achernar} suggest that there ought to be
many single-mode Be stars.  They would have to function somewhat
differently but with the same basic outcome, namely a decretion disk,
and an explanation is needed for them.

For instance, if mass loss opens a valve through which pulsation
energy leaks out, the interaction between two NRP modes opens and
closes this valve periodically.  But in a single-mode pulsator it
would stay open until the energy supply is temporarily exhausted and
needs to be re-built.  There is also the suggestion by
\cite{1986A&A...163...97A} that nonradial $g$-modes can act as stellar
core-to-surface carriers of angular momentum \citep[see also
][]{2014MNRAS.443.1515L}.

Alternatively, apparent single-mode pulsators with outburst repetition
timescales of a decade like 28 $\omega$ CMa and Achernar may actually
be multimode pulsators with frequency separations of as little as
3\,10$^{-4}$\,c/d.  Existing ground-based observations do not have the
necessary precision, and space data do not have enough time coverage,
to reject or confirm such a hypothesis so that any new observational
effort would require a long shot well into the next decade.

\subsubsection{The outer engine:  Large-scale gas circulation flows}

The outer engine is not any less difficult to reverse-engineer because
stars other than Be stars do not cast the matter they lose into a
Keplerian disk and so cannot provide much guidance.  Therefore, the
following attempts to bootstrap the properties of the outer engine
that regulates this process:
\begin{list}{$\bullet$}{\partopsep=0mm\topsep=0mm\itemsep=0mm} 

\item
The outer engine is located in the transition region between star and
disk because the {\v S}tefl frequencies are associated with
superequatorial velocities.  

\item 
It is fed by the inner engine,
which is powered by low-order nonradial $g$-mode pulsation as
discussed in Sect.\,\ref{NRPinteract}.  

\item 
The outer engine stops
working when the inner engine (pulsation) does not deliver enough
matter (and associated energy and angular momentum) to it.  

\item
Because of the appearance of the {\v S}tefl frequencies in spectral
lines, the outer engine seems to be tied to the azimuthal modulation
of the gas density just above the photosphere proper.  

\item 
The {\v
S}tefl frequency of \object{$\mu$ Cen} manifests itself in the
spectroscopy but not in the photometry.  In $\eta$ Cen, the {\v S}tefl
frequency is revealed by both observing techniques.  \object{$\mu$
Cen} is seen at a small inclination angle while \object{$\eta$ Cen} is
viewed through its disk.  Because the presumed outer engine is close,
but not very close, to the star, this could mean that in
\object{$\eta$ Cen} there is an occultation process at work, and in
\object{$\mu$ Cen} the outer engine is visible all the time.  

\item
The operating frequency of the outer engine (the {\v S}tefl frequency)
is slightly lower than the frequency of the related pulsation.  

\item
This could imply that the azimuthal period of the density modulation
is about the same as the period of the nonradial quadrupole modes 
typical of Be stars \citep{2003A&A...411..229R}, namely 180 degrees. 

\item 
The operating frequency is not phase
coherent but wobbles at the few-percent level.  

\item 
As a result, the
hypothesized regions of enhanced density do not have a stable
azimuthal position but can drift by significant amounts.  

\item 
This
would disfavor stellar magnetic fields as the ``anchor'' of the
density enhancements unless such magnetic fields also drift around.
This is in agreement with the magnetic-field measurements mentioned
above.  

\item 
Maybe a simpler mechanism for the density modulation is dynamic:
For instance, there could be large-scale circulation flows and the
density be enhanced where the circulation velocities are lower or
cause photometric variations in some other way.  That is, the observed
circumstellar variations are no caused fixed lumps of orbiting matter
but by pile-ups in the flow.

There are at least two ways by which this could work (if at all):
\begin{list}{$\circ$}{\partopsep=0mm\topsep=0mm\itemsep=0mm}
\item
At the inner edge of gaseous Keplerian disks, there can be circulatory
motions known as Rossby Wave Instability \citep{1999ApJ...513..805L, 
2000ApJ...533.1023L}.  Perhaps, they could also provide
the seed for the viscosity of the decretion disks of Be stars.  

\item

In the terrestrial atmosphere and all gaseous planets of the solar
system, there are circulatory gas flows, which are related to Rossby
waves. Of particular interest is possibly that these gas flows, which
are driven by fast rotation and temperature differences, have links to
jet streams.  See \cite{Oishi} for the Earth,
\cite{2012Icar..218..817S} for Jupiter, \cite{2012Icar..219..689D} for
Saturn, and \cite{2013Natur.497..344K} for Uranus and Neptune.

That is, these are powerful, energetic processes.  But note that not
even in the solar system is the formation of jet streams fully
understood \citep{2007Sci.315..467K}.

\end{list} 

\item 

In Fourier space, a {\v S}tefl frequency, which in \object{$\eta$ Cen} is
sitting amidst the Type II frequency group at 1.55\,c/d, would only be
the tip of an iceberg consisting of all the intrinsic noise and
nonlinearities associated with the circulation process.  Group
harmonics will arise very naturally.  Irrespective of whether the
circulation motions are in the outer stellar atmosphere or the inner
disk, still images and movies of planetary atmospheres may provide some
guidance as to what to search for.  This is a very close match of the
model-free description given above of the photometric Type II
frequency groups in Be stars.

\end{list}

\noindent 
The innermost part of the disk would thus consist of spatially
quasi-periodic gas circulation cells.  They would owe their existence
to one or more of (i) the rapid rotation of the central star, (ii) the
large pole-to-equator temperature differences, (iii) the detachment
process of the gas leaving the star (and, in Be stars, partly
returning to it from a viscous disk), and (iv) the variation of the
local mass-loss rate with stellar azimuth.  The gas temporarily
trapped in these cells could be supplied by stellar $g$-mode pulsation
when the amplitude is large enough.

The above design of the outer engine was derived in a purely empirical
way. It does not compete with the VDD model but may rather supply it
with inner boundary conditions.  It also has major implications for
the interface between inner and out engine.

\subsubsection{Coupling two engines running at very different frequencies}

Describing this interface is, in fact, a challenge:  Any model for the
mass loss from Be stars must explain how, on the one hand, the
variabilities with the NRP (to be precise: the slow process linked to
the pulsation) and {\v S}tefl frequencies can be so closely coupled
and how, on the other hand, such a causal chain does not get broken by
the difference in the frequency.  The difference in observed frequency
may be related to the phase velocity of the NRP waves. However, not
even the sign of the latter is clear: \cite{2003A&A...411..229R} found
retrograde modes to well describe their spectroscopic observations of
two dozen Be stars.  By contrast, theoretical models often prefer
prograde modes, especially if they are `tasked' with supporting the
star-to-disk angular-momentum transfer; see \cite{2014arXiv1412.8511K}
for a discussion.

But the frequency difference exists, and so does the mystery of two
processes that are closely connected but operate at quite different
frequencies.  The mystery is probably extended by the fact that the
{\v S}tefl frequencies seem to `track' the pulsation frequencies over
a range of at least a factor of $>2$ when all stars with
available data are considered together (see Table \ref{freqtab}).  The
concept of large-scale circulation motions can also deliver on this
challenge.  Because they seem to prevail only during phases of
significant mass loss that feeds them, the matter participating in
these motions is permanently replaced with new one.  Therefore, the
observed frequencies of these structures are kind of phase frequencies
whereas the physical causality is maintained by the physical motions
of the gas.  In this way, a quadrupole mode can supply two huge eddies
with matter, which nevertheless propagate with a different azimuthal
velocity.

In the proposed context, it does not appear surprising that the
amplitude of the {\v S}tefl frequency scales with residuals from the
mean brightness: A larger stellar action (mass loss) leads to a larger
circumstellar response.  The variable phase relation suggests that the
logical rope linking the two processes must be fairly elastic.  The
approximate anti-correlation between the {\v S}tefl frequency and its
amplitude can be understood such that when the mass loss is higher,
there is more matter close to the star, where orbital frequencies are
higher.  Such a change in frequency is nothing but a change in phase
so that the variable phase is also explained.

\subsection{Other early-type stars} 

The above description places the circulation streams in the inner
disk.  But the available observations cannot rule out similar motions
in the outer stellar atmosphere.

If there is an analogy between Be stars and the atmospheres of rocky
as well as rapidly rotating gaseous planets, these cases would
probably bracket other stars with significant rotation and extended
atmospheres.  The phenomenon that such a conjecture brings immediately
to mind are the Discrete Absorption Components (DAC's) in the UV wind
lines of virtually all massive stars.  \cite{2015ApJ...809...12M} have
recently shown that they must arise from very close to the
photosphere.  At 15-20\% of the stellar diameter, their footprints
are quite large. 

Radiatively driven winds are so extremely intrinsically unstable that
a considerable variety of large-scale perturbations seem, in
principle, capable of growing into DAC's
\citep{1996ApJ...462..469C}. The challenge is to explain the apparent
semi-regular angular partition.  Based on the analysis of 10 O-type
stars, \cite{1999A&A...344..231K} argue that the azimuthal period is
often (but clearly not always) close to 180 degrees.  Magnetic fields
might be too rigid, given the frequent large deviations from genuine
periodicity.  More importantly, MiMeS has found global magnetic fields
in less than 10\% of all early-type stars \citep{2014IAUS..302..265W}.
In a targeted search in 13 OB stars with prototypical DAC behaviour,
\cite{2014MNRAS.444..429D} did not detect dipolar magnetic fields in
any of them.  But DAC's occur in virtually all luminous O stars
\citep{1989ApJS...69..527H}.  Nonradial pulsations are perhaps
similarly ubiquitous among stars with radiative winds so that their
native angular periodicity makes them a stronger contender for the
explanation of DAC's than magnetic fields.  However, low-order NRP's
do not seem to be common in O-type stars (see, e.g., the examples of
$\zeta$ Oph, $\xi$ Per, and $\zeta$ Pup mentioned below).

Gas circulations in the transition region between photosphere and base
of the wind might constitute a plausible alternative.  But the mostly
much smaller temperature contrast between equator and poles than in Be
stars, in which it is caused by the fast rotation and may even lead to
significant equatorial convective motions, may not be able to channel
enough energy into this process.  Moreover, the much larger radiation
pressure could inhibit such flows altogether.

Recently, \cite{2014MNRAS.441..910R} reported on observations with the
{\it MOST} satellite of the bright O star $\xi$ Per (O7.5
III(n)((f))).  There are several 1-mmag variations near 0.5 c/d, which
is interpreted as twice the rotation frequency.  Assuming that there
are two roughly identical regions along the stellar circumference,
\cite{2014MNRAS.441..910R} attribute the photometric variability to
photospheric spots.  The best description found of the data is that
the spots propagate at one and the same angular rate (that of the
stellar rotation) but are relatively short-lived with spots
disappearing after a while and new ones developing.  If the size of
these spots is as large as inferred by \cite{2015ApJ...809...12M},
their brightness contrast must be quite low.

It seems difficult to distinguish this photometric variability of
$\xi$ Per from the {\v S}tefl frequencies of $\eta$ and \object{$\mu$ Cen} and
\object{$\alpha$ Eri}: low amplitude, roughly integer fraction of the
rotational timescale (with rotational eclipses), and poor phase
coherence.  If this means that the phenomena are of the same nature,
and if gas circulation cells are responsible for them, they would in
$\xi$ Per reside in the star because Be-star disks are neither known
nor expected to develop around very hot and luminous stars.  In
agreement with this, $\xi$ Per does not exhibit the very high noise
associated with {\v S}tefl frequencies in Be stars.  Another
difference is that O stars do not seem to pulsate in low-order
nonradial modes (and the higher-order pulsation does not appear to
modulate the mass loss).

Another bright prototypical O star is \object{$\zeta$ Pup} (O4
If(n)p), albeit with above-average space and rotational velocities.
BRITE has observed this star, too.  Ramiaramanantsoa et al.\,(in
preparation) find that this star's behaviour is similar to that of
$\xi$ Per but more stable.  The frequency of 0.56\,c/d was also seen
for nearly four years by SMEI \citep{2014MNRAS.445.2878H} and falls
into a grey zone between rotation and pulsation.  The proximity to the
Sun of $\xi$ Per and \object{$\zeta$ Pup} attracts the same argument
as for $\mu$ (and now also $\eta$) Cen: There must be many more such
stars.

If the facts put together in this subsection do belong together, any
attempt of their explanation is faced with the combination of large
spatial size, small photometric contrast, and high significance in
velocity.  Circulation flows seem capable of coping with such a
challenge. In a study that has remained without follow-up by the
hot-star community, \cite{1990NYASA.617..190D} have undertaken a first
theoretical reconnaissance investigation of the properties of putative
large-scale vorticities in rapidly rotating hot-star atmospheres.
They conclude that the associated photometric contrast will be low
because the vorticities are shallow and so have little
flux-(de)focusing power. Dowling \& Spiegel also consider that such
vorticities may be related to the ubiquitous microturbulence invoked
to explain excess line broadening.  This is quite similar to the
proposal by \cite{1976ApJ...206..499L}, which, however, is based on
multi-mode nonradial pulsation.  The two explanations can probably
co-exist \citep[also with the convection-based model of
][]{2009A&A...499..279C}.  But the empirical evidence of nonradial
pulsation in early-type stars is mostly limited to few discrete modes
(cf.\,$\xi$ Per and $\zeta$ Pup). Vorticities may more readily lead to
a continuum of motions.

\subsection{Options for verification} 

Because of their velocity contrast, the hypothesized circulation
regions in Be stars would probably have a small radial extent, just
some fraction of the stellar radius.  Such considerably
sub-milliarcsec structures are too small for current optical and
infrared interferometers, which, for observations of Be-star disks,
often use the central stars as unresolved calibrators.

In shell stars, where the line of sight passes through the plane of
the disk, the disk structure can also be probed by high-resolution
spectroscopy.  There is a seemingly unparalleled series of
high-resolution spectra by \cite{1996A&A...312L..17H} of \object{48
Lib} (B3\,IVe-sh), which recorded the presence for a few months as
well as the very rapid variability of multiple narrow absorption
components in optical low-excitation shell lines.  Because of its
uniqueness it has not found an understanding based on a broad interest
and discussion.  Hanuschik \& Vrancken offer as one explanation
`higher-order, local distortions in the global spiral pattern'
\citep[referring to global disk oscillations as described
by][]{1991PASJ...43...75O}.  It could just as well be the low-spatial
frequency noise farther out in the disk of the circulation flows.
That is, spectroscopy at high time and spectral resolution of shell
stars could contribute to unraveling the {\v S}tefl process.

Among the available observables, polarization is probably the one most
strongly focused on the conditions in the immediate vicinity of the
central star as has been shown by \cite {1984ApJ...279..721H},
\cite{1984ApJ...287L..39G}, and \cite{2007ApJ...671L..49C}.  It could
be useful to try to compare temporal power spectra of fast polarimetry
with spatial power spectra of models of near-stellar disk structures.

{\v S}tefl frequencies were discovered in series of spectra of Be
stars without strong shell-star characteristics.  However, while the
{\sc Heros} spectra \citep{1998ASPC..135..348S} were sufficient for
detection purposes, the detailed nature of the variability cannot be
derived from them.  The separation in radial velocity of the {\v
S}tefl variability from much of the photospheric variability is an
asset but the contamination will still be large.  This would make a
new spectroscopic campaign quite expensive with uncertain outcome.

In photometric power spectra, the separation of {\v S}tefl and main
pulsation frequencies is more favorable.  In long time series, the
width of discrete pulsation frequencies will be negligible for the
given purpose.  If in addition the sampling is high, power spectra can
be calculated in sliding time windows.  This will unambiguously
establish whether the `grass' in the power spectra of Be stars is or
is not phase coherent, i.e., whether it is rooted deep inside the star
or grows in the star-disk transition region.  The four-year {\it
Kepler} database satisfies these requirements and forms the basis 
of Paper II, which also deploys wavelet analyses.  

This reasoning ignores the possibility of stochastically excited
pulsation modes \cite[e.g.,][]{2012A&A...546A..47N}.  It would be very
valuable to secure spectra of the B-type stars observed by {\it
Kepler} and other satellites that exhibit power spectra with
particular rich groups of transient spikes.  If stars with such
properties are mainly Be stars, the non-stellar origin of Type II
frequency groups would be further strengthened.  Such efforts should
take into account the widespread volatility of Be-star disks and the
emission lines forming in them.

One of the most promising follow-up projects should be the comparison
of (equator-on) Be stars to very rapidly rotating B-type stars without
disks and emission lines, so-called Bn stars.  Thirty years ago,
\cite{1986PASP...98...35P} reported that only Be stars show low-order
NRP's while Bn stars -- and many Be stars -- pulsate in high
nonradial-order modes.  A possibly related observation
\citep{1990ASIC..316..235W} is that with increasing rotational
velocity the prevalent nonradial mode order, $m$, may also increase.
\cite{1993A&A...273..135A} pointed out that this may only be apparent
because at higher rotation rates larger toroidal terms alter the
velocity field of low-order modes.  However, a comparison of Bn and
equator-on Be stars may be less susceptible to such effects.
Moreover, SPB stars have been added to the B-type stars with
long-period, low-order $g$-modes, and some of Them show relatively
rapid, but not extreme, rotation \citep[e.g.,][]{2015MNRAS.446.1438D}.
The triangle of Be / Bn / SPB stars should provide for valuable
intercomparisons.  Moreover, Bn and SPB stars should be free of
circumstellar noise and so indirectly, by elimination, contribute to
understanding the latter.

\section{Conclusions} 

The good agreement between BRITE photometry, {\sc Heros} spectroscopy,
and other observations taken up to more than two decades apart
suggests that, for the combinability of observations of Be stars, high
Fourier quality (sampling, resolution) is at least as important as
proximity in time.  This consistency deserves to be contrasted with
the still widespread belief that Be stars behave in an erratic way.
The latter is quite certainly true if their sometimes-complex
variability is not properly captured by the observations, be it
because of too sparse sampling or too short data strings or both.  But
insufficient data is not an intrinsic property of Be stars.

Thanks to the so enabled combination of BRITE photometry and {\sc
Heros} spectroscopy it is possbile to take advantage of the fact that
nature has been benign in orthogonalizing the Be problem: It can be
separated into an inner and an outer engine.  In addition to
confirming this, the new BRITE data provide a more detailed view of
the complex interface between the inner and outer Be-star engines.  The
{\v S}tefl frequencies seem to be the key element of this interface.

Contrary to stellar pulsations, the circumstellar {\v S}tefl
frequencies exhibit substantial `deficiencies' in phase coherence and
are accompanied by numerous other similarly deficient frequencies.  If
they are mistaken for stellar pulsations, they may be (incorrectly)
interpreted as a large number of pulsation frequencies.  The
BRITE observations of $\eta$ and \object{$\mu$ Cen} suggest that the
complexity of the power spectra of Be stars does not lie in exotic
variations of stellar pulsations that, if related to $g$-modes, are
rooted deeply inside the star but rather in temporary circumstellar
processes that are only quasi-periodic.

At the most elementary level, the variability of Be stars like $\eta$
and \object{$\mu$ Cen} may possibly be characterized by as few as
three numbers: Two stellar pulsation frequencies and a slightly lower
circumstellar {\v S}tefl frequency.  In addition, a formula is needed,
which describes how the two NRP modes join (or even amplify) their
forces.  The {\v S}tefl frequency is accompanied by a Type II
frequency group and 2:1 group harmonics.  The sometimes reported
extreme multi-periodicity of Be stars during outbursts would be an
illusion due to the noise of the circumstellar activity, for instance,
in large gas-circulation flows.  Its diagnostic value would not be
diminished, though, but shift from asteroseismology to the
hydrodynamics of the mass-loss process.  The variations of the {\v
S}tefl frequencies in value and amplitude seem to be good tracers of
the mass supply from the star to the disk.

A still simpler variant would not require an interaction of two
short-term variations.  A single NRP mode energizes the outer
mass-loss engine directly and supplies it with angular momentum from
deep inside the star until its own energy supply is exhausted after
some years.

The distinction made between multi- and single-mode pulsators among Be
stars does not lead to a black-and-white picture: Multi-mode pulsators
like $\zeta$ Oph (O9.2 IVe) may show variable pulsation amplitudes and
undergo outbursts but do not exhibit any obvious signs of outbursts
being related to some combination of the pulsations
\citep{2014MNRAS.440.1674H}.  However, the frequencies of up to
7.2\,c/d are mostly on the high side for low-order $g$-modes.
Spectroscopic evidence is available for only some of the variations
and indicative of medium $m$ values.  That is, $\zeta$ Oph does not
contradict the interacting-modes paradigm but may have more in common
with the single-mode pulsators.

A valid concern about the rotation-plus-pulsation hypothesis for the
mass-loss mechanism from Be stars can be deduced from the apparent
dichotomy between early-type Be stars, which are invariably quite
active, and Be stars of late spectral sub-classes, which often hardly
display any activity at all \citep{2013A&ARv..21...69R}.  This may be
just a difference in amplitude: Star \object{KIC\,11971405} is of
spectral type B8 but the precision photometry with {\it Kepler}
\citep{2015MNRAS.450.3015K} shows it to be pulsating and probably
undergoing mass-loss events.  Pulsations were also detected by BRITE
in other late-type Be stars (Baade et al., in prep.).

If the coupling of NRP modes can release more energy than their simple
linear combination, it could enable the formation of disks also around
late-type Be stars with low NRP amplitudes.  This could be helped by a
fractional critical rotation that increases with decreasing mass
\citep[see][for a detailed
discussion]{2013A&ARv..21...69R}. Furthermore,
\cite{2015MNRAS.454.2107V} find that disks around late-type Be stars
have much lower densities.

In the atmospheres of solar-system planets, jets are powered by
rotation and temperature differences, in accretion disks Rossby waves
with similar circulatory motions are predicted
\citep{1999ApJ...513..805L, 2000ApJ...533.1023L}, and
\cite{1990NYASA.617..190D} discuss jets for hot-star atmospheres.
That is, these circulation flows are significant power houses and
could form part of the outer engine.  If they do occur, their
high-momentum end could also provide the physical basis for the
mass-ejection process in the model of \cite{1997ASPC..121..494K},
which otherwise laid a promising foundation to a realistic
reproduction of the disk variability of \object{$\mu$ Cen} following
outbursts.

In summary, a cartoon inspired by \object{$\eta$ Cen} of a Be star in outburst 
could look as follows:
\begin{list}{$\bullet$}{\itemsep=0mm\topsep=0mm}
\item
A double wave is propagating along the stellar equator.  It
results from the temporary
superposition of two or more nonradial quadrupole modes. 
\item
Nonlinear coupling at the difference frequency of the NRP modes 
leads to significant amplification beyond the 
amplitude sum of the parent frequencies.  
\item
At phases of maximal amplitude, outbursts are triggered.  
\item
The matter so propelled flows to, and is
circulated through, two large eddies at the base of the disk.  These
eddies propagate at a different angular velocity than the stellar
pulsation waves.  It is the sum of the local Keplerian angular
velocity and any phase velocity of the vortices and corresponds to the
{\v S}tefl frequency.  The latter is not constant because the vortices 
may move both radially and azimuthally.  

\item
Conservation of the total angular momentum limits the fraction of the
matter that does reach orbital velocities to a small number.  The rest
loses angular momentum in a plethora of small eddies and eventually
falls back to the star \citep[cf.\,][]{2002MNRAS.337..967O,
2013A&A...559L...4R}.  These small circulation cells are the source of
the noise seen during outbursts and may be the seed of the viscosity
in the VDD model.  It may be fitting that a first determination of the
viscosity parameter in a Be disk yielded a rather high value
\citep{2012ApJ...744L..15C}.

\end{list}

\noindent Forthcoming papers in this BRITE string of the series will
deal with late-type Be stars and a comparison of Be and Bn stars.
Among the studies of individual Be stars, there will be one of 27 and
28 $\omega$ CMa (which are being observed by BRITE while this paper is
being written).  They form a (nonphysical) couple of early-type Be
stars very similar to $\eta$ and \object{$\mu$ Cen} with one shell
(27) and one high-inclination (28 $\omega$) star.  In neither of them,
multiple frequencies have been detected to date.

\begin{acknowledgements} 

DB and TR dedicate this paper in gratitude to their long-term
collaborator and friend Stanislav (Stan) {\v S}tefl (1955 - 2014),
with whom this project had been planned like so many others before.
The authors thank the BRITE operations staff for their untiring
efforts to deliver data of the quality that enabled this
investigation.  They appreciate the permissions by Tahina
Ramiaramanantsoa et al.\ to mention some early results of their
ongoing work on \object{$\zeta$ Pup}.  Coralie Neiner is thanked for
the early communication of magnetic-field measurements with ESPaDOnS
of $\eta$ Cen.  This research has made use of the SIMBAD database,
operated at CDS, Strasbourg, France.  This research has made use of
NASA's Astrophysics Data System. ACC acknowledges support from CNPq
(grant number 307076/2012-1) and FAPESP (grant number 2015/16592-0).
GH thanks the Polish NCN for support (grant 2011/01/B/ST9/05448).
AFJM is grateful for financial aid from NSERC (Canada) and FRQNT
(Quebec).  APi acknowledges support from the Polish NCN grant
no. 2011/03/B/ST9/02667.  APo acknowledges support through NCN grant
No.\ 2013/11/N/ST6/03051.  GAW acknowledges Discovery Grant support
from the Natural Sciences and Engineering Research Council (NSERC) of
Canada.  The authors from Poland acknowledge assistance by BRITE PMN
grant 2011/01/M/ST9/05914.

\end{acknowledgements}

\bibliography{dbaade}

\end{document}